\documentclass[twocolumn,tighten,numberedappendix,twocolappendix]{aastex631}
\usepackage{amsmath}

\defcitealias{Beloborodov22}{B22}

\def\simlt{\lower.5ex\hbox{$\; \buildrel < \over \sim \;$}}
\def\simgt{\lower.5ex\hbox{$\; \buildrel > \over \sim \;$}}

\def\beq{\begin{equation}}
\def\eeq{\end{equation}}
\def\ba{\begin{eqnarray}}
\def\ea{\end{eqnarray}}
\def\bB{\boldsymbol{B}}
\def\bE{\boldsymbol{E}}

\def\Eq{Equation}
\def\Eqs{Equations}

\def\sT{\sigma_{\rm T}}

\def\Gw{\Gamma_{\rm w}}
\def\Lw{L_{\rm w}}

\def\sigw{\sigma_{\rm w}}

\def\E{{\cal E}}
\def\N{{\cal N}}
\def\dN{\dot{\cal N}}

\def\RLC{R_{\rm LC}}

\def\tom{\tilde{\omega}}

\def\tC{t_{\rm C}}

\def\tn{\tilde{n}}

\def\tom{\tilde{\omega}_p}
\def\tn{\tilde{n}}
\def\vgr{v_{\rm gr}}

\def\omB{\omega_B}

\def\Bbg{B_{\rm bg}}
\def\bBbg{\boldsymbol{B}_{\rm bg}}
\def\Ebg{E_{\rm bg}}
\def\bEbg{\boldsymbol{E}_{\rm bg}}

\def\xirise{\xi_{\rm rise}}

\def\tBbg{\tilde{B}_{\rm bg}}
\def\tomB{\tilde{\omega}_B}
\def\tB{\tilde{B}}
\def\tE{\tilde{E}}

\def\P{{\cal P}}

\def\gav{\langle\gamma\rangle}
\def\tgav{\langle\tilde\gamma\rangle}
\def\uzav{\langle u_z\rangle}
\def\tuzav{\langle \tilde{u}_z\rangle}
\def\bb{\boldsymbol{\beta}}

\def\gD{\gamma_{\rm D}}
\def\bD{\beta_{\rm D}}
\def\vD{v_{\rm D}}
\def\bvD{\boldsymbol{v}_{\rm D}}

\def\bs{\beta_{\rm s}}

\def\trho{\tilde{\rho}}

\def\trhou{\trho_{\rm u}}

\def\c{\kappa}
\def\cu{\c_{\rm u}}

\def\uzav{\langle u_z \rangle}

\def\P{{\cal P}}

\def\Tp{T_{\rm p}}

\def\Pp{P_{\rm p}}
\def\Up{U_{\rm p}}

\def\Bu{B_{\rm u}}
\def\gu{\gamma_{\rm u}}

\def\betau{\beta_{\rm u}}

\def\tcross{t_{\rm cross}}

\def\sigu{\sigma_{\rm u}}

\def\R{{\cal R}}

\def\me{m}

\def\Uw{U}

\def\KF{{\cal K}}
\def\tKF{\tilde{\cal K}}

\def\bu{\boldsymbol{u}}

\def\tu{\tilde{u}}
\def\tg{\tilde{\gamma}}
\def\dEe{\dot{\E}_e}
\def\P{{\cal P}}
\def\dPe{\dot{\P}_e}

\def\tomL{\tilde{\omega}_{\rm L}}

\def\tbeta{\tilde{\beta}}
\def\tU{\tilde{U}}
\def\gD{\gamma_{\rm D}}

\def\tbD{\tilde{\beta}_{\rm D}}

\def\cs{\kappa_{\star}}

\def\om{\omega}
\def\L{L}
\def\tw{T}
\def\rhou{\rho_{\rm u}}
\def\bBu{\bB_{\rm u}}

\def\Uf{U_{\rm f}}
\def\Pf{P_{\rm f}}
\def\tom{\tilde{\om}}

\def\tUw{\tilde{U}}
\def\uD{u_{\rm D}}

\def\trel{t_\star}
\def\rhost{\rho_\star}

\def\tC{\tilde{C}}
\def\bEw{\bE_{\rm wave}}
\def\Ew{E_{\rm wave}}
\def\bBw{\boldsymbol{B}_{\rm wave}}
\def\Bw{B_{\rm wave}}
\def\PMHD{\P_{\rm MHD}}
\def\Psc{\P_{\rm sc}}
\def\R{{\cal R}}
\def\Cstoch{C_{\rm stoch}}
\def\Ctrap{C_{\rm trap}}
\def\tomBu{\tomB^{\rm u}}
\def\tomu{\tom_{\rm u}}
\def\rstoch{r_{\rm stoch}}

\def\Lesc{L_{\rm esc}}
\def\Q{{\cal Q}}
\def\gtrap{\gamma_{\rm trap}}

\def\tgrad{\tg_{\rm rad}}
\def\Eplasma{{\cal E}_{\rm p}}
\def\Emag{{\cal E}_{\rm mag}}
\def\Erad{{\cal E}_{\rm rad}}

\def\tgstoch{\tg_{\rm stoch}}
\def\aw{a_{\rm wave}}
\def\rhoav{\langle\rho\rangle}
\def\trhoav{\langle\trho\rangle}
\def\ph{\varphi}
\def\tdge{\tilde{\dot\gamma}_e}

\def\rb{r_b}

\def\astoch{a_{\rm stoch}}

\def\bs{b_{\rm s}}
\def\bup{b_{\rm u}}

\newbox\grsign \setbox\grsign=\hbox{$>$} \newdimen\grdimen \grdimen=\ht\grsign
\newbox\simlessbox \newbox\simgreatbox \newbox\simpropbox
\setbox\simgreatbox=\hbox{\raise.5ex\hbox{$>$}\llap
     {\lower.5ex\hbox{$\sim$}}}\ht1=\grdimen\dp1=0pt
\setbox\simlessbox=\hbox{\raise.5ex\hbox{$<$}\llap
     {\lower.5ex\hbox{$\sim$}}}\ht2=\grdimen\dp2=0pt
\setbox\simpropbox=\hbox{\raise.5ex\hbox{$\propto$}\llap
     {\lower.5ex\hbox{$\sim$}}}\ht2=\grdimen\dp2=0pt
\def\simgt{\mathrel{\copy\simgreatbox}}
\def\simlt{\mathrel{\copy\simlessbox}}

\begin{document}

\title{Compression fronts from fast radio bursts}

\email{amb@phys.columbia.edu}

\author[0000-0001-5660-3175]{Andrei M. Beloborodov}
\affiliation{Physics Department and Columbia Astrophysics Laboratory, Columbia University, 538  West 120th Street New York, NY 10027,USA}
\affil{Max Planck Institute for Astrophysics, Karl-Schwarzschild-Str. 1, D-85741, Garching, Germany}

\begin{abstract}
When a fast radio burst (FRB) expands from its source through a surrounding tenuous plasma, it strongly heats and compresses the plasma at radii up to $\sim 10^{14}$\,cm. The likely central engines of FRBs are magnetars, and their ambient plasma at radii $r\gg 10^{10}$\,cm is a magnetized $e^\pm$ wind. We formulate basic equations of the FRB-plasma interaction, solve them numerically, and describe the physical picture of the interaction. We find the following: (1) FRBs emitted at $r<\rstoch\sim 10^{12}$\,cm induce fast stochastic heating and strong compression of the wind, sweeping it like a broom. The outcome of this interaction is determined by the energy losses of the radio wave; we evaluate the parameter space where FRBs survive and escape. (2) At radii $r>\rstoch$, FRB induces regular particle oscillations in the radio wave with the standard strength parameter $a$, and drives a compression wave in the wind. At $r>r_\star\sim 10^{13}\,$cm, the compression wave becomes locally quasisteady, with compression factor $1+a^2$. FRBs avoid damping if they are released into the wind medium outside $r_{\rm damp}\sim 10^{11}$\,cm.
\end{abstract}



\section{Introduction}

Observations of fast radio bursts (FRBs) demonstrate the existence of ultrastrong electromagnetic waves in the Universe. They have characteristic frequencies $\nu=\om/2\pi\sim 1$\,GHz and durations $T\sim 1$\,ms \citep{Petroff19}, and their dimensionless strength parameter is\footnote{We use the standard notation $Q_n$ for a quantity $Q$ normalized to $10^n$ in CGS units, e.g. $\nu_9=\nu/10^9 {\rm Hz}$ and $r_{12}=r/10^{12}{\rm cm}$.}
\beq
\label{eq:a}
   a \equiv \frac{e\langle\Ew^2\rangle^{1/2}}{\me c\, \om} \approx 16\,\nu_9^{-1}\L_{42}^{1/2}r_{12}^{-1},
\eeq
where $\me$ is the electron mass, $c$ is the speed of light, $\L=c r^2 \langle \Ew^2 \rangle$ is the isotropic equivalent of wave luminosity; $\langle \Ew^2 \rangle^{1/2}$ is the root-mean-square of the wave electric field when the wave expands to a radius $r$.

FRBs have to propagate through the ambient plasma surrounding the source.
The likely central engines of FRBs are neutron stars, which are surrounded by electron-positron plasma. The plasma is strongly magnetized: its magnetic stress $B^2/4\pi$ far exceeds the rest-mass energy density $\rho c^2$,
\beq
  \sigma\equiv\frac{B^2}{4\pi \rho c^2}\gg 1,
\eeq
where $B$ and $\rho$ are measured in the plasma rest frame. Near the neutron star, the plasma is locked in the closed magnetosphere, rotating with a period $P\sim 1-10$\,s. Outside the light cylinder $\RLC=cP/2\pi\approx 5\times 10^{9}(P/1{\rm \,s})$\,cm, a relativistic wind flows from the magnetosphere, powered by the star's rotation. Both the magnetosphere and the wind have $\sigma\gg 1$ (see \cite{Cerutti17} for a review).

The plasma around neutron stars is tenuous, however it can have a strong effect on FRB propagation. The radio wave with $a\gg 1$ accelerates plasma particles to high energies, and this process can damp the wave, in particular if the energized plasma experiences radiative losses. Strong radiative damping has been demonstrated for FRBs emitted near a magnetar and propagating through its equilibrium outer magnetosphere at radii $r\sim 10^8-10^9$\,cm \citep{Beloborodov21b,Beloborodov24}. 
In addition, strong radio waves are damped through excitation of small-scale Alfv\'en waves in the magnetosphere \citep{Golbraikh23}.

Recently, \cite{Sobacchi24} suggested that FRBs may also be damped at larger radii $r\gg\RLC$ by the interaction with the wind. They argued that damping results from compression of the magnetized wind by the FRB, efficiently consuming FRB energy if it is emitted at $r\simlt 10^{12}$\,cm. They calculated compression for a nearly monochromatic FRB using an expansion in Fourier harmonics, and also used two assumptions: (1) particles perform regular oscillations in the radio wave, and (2) the pattern of plasma motion in the wave packet is steady in the frame moving with the packet. As shown below, 
both assumptions hold at $r\gtrsim 10^{13}$\,cm and should not be used when analyzing FRB propagation at smaller radii.

The present paper investigates the FRB-wind interaction using a new method. The problem is stated in section~\ref{formulation} and then solved in two steps:

First, we describe plasma response to an FRB on a ``micro'' level by tracking the motion of individual particles (section~\ref{heating}). The response can be regular oscillation or stochastic heating  (\citealt{Beloborodov22}; hereafter \citetalias{Beloborodov22}). It is best viewed in the local center-of-momentum frame of the plasma particles (``fluid frame''). Remarkably, the microscopic response can be described independently of the fluid motion. This fact is illustrated with simulations of particles driven by FRBs in static and accelerating magnetized fluids. The simulations demonstrate the regular and stochastic response, and  the transition between the two regimes.

Then, we examine the macroscopic fluid dynamics of plasma exposed to FRBs. The magnetized fluid is pushed and compressed by the FRB on timescales much longer than the particle Larmor period and the wave oscillation period,\footnote{The acceleration effect may be thought of as a ponderomotive force exerted by the radio wave. In the presence of strong radiative losses it is more suitable to view it as a scattering process. In any case, it may be described from first principles by solving for particle motion in the wave and then examining the fluid dynamics on macro scales, as done in this paper.}
so it obeys magnetohydrodynamic (MHD) equations. The effective enthalpy of the MHD fluid is determined by particle motions in the fluid frame (i.e. by the microscopic response to the FRB) and controlled by the Lorentz-invariant parameter $a$. One can think of the micro-level response as a local heating process and the macro-level response as bulk acceleration of the fluid, leading to its compression. Equations for fluid dynamics inside a propagating radio wave packet are formulated in section~\ref{MHD}. 

It is easy to see why plasma heating by FRBs is inevitably accompanied by its bulk acceleration: any energy $\Delta\E$ received from the radio wave comes with momentum $\Delta\E/c$ directed along the wave propagation direction (the wave loses momentum $\Delta\E/c$). \cite{Beloborodov21b} estimated this effect for FRBs propagating through the dipole magnetosphere of a neutron star, where the FRB-plasma interaction is highly radiative and can be described as a scattering process. A more general and rigorous calculation should be formulated as a time-dependent fluid dynamics problem, using MHD equations and allowing weak or strong radiative losses of the heated plasma. This is attempted in the present paper. The weakly radiative regime becomes particularly relevant when the FRB propagates through the wind far outside the closed magnetosphere.

While heating by FRBs implies push and compression, the compression in turn can change the heating process. Note that the MHD fluid is compressed together with its magnetic field. Once the magnetic field exceeds a certain value, regular particle oscillations in the radio wave change into stochastic heating, as shown below. This change boosts the heating rate and leads to even stronger compression, with enhanced radiative losses of the heated particles. The compression fronts driven by FRBs in magnetar winds involve quite dramatic effects, capable of damping the FRB itself as discussed in later sections of this paper.

Before considering FRBs expanding through a radial wind, we examine the compression front in a plane geometry with a uniform background. First, we show that the problem has a simple steady-state solution (section~\ref{steady}), which agrees with the results of \cite{Sobacchi24}. Its compression factor $C_\star$ is directly related to the particle energy gained from the radio wave and becomes huge for waves with strength parameter $a\gg 1$. Remarkably, $C_\star$ is independent of the plasma magnetization $\sigma$. This fact casts doubt on the relevance of the steady-state solution; in particular, it clearly becomes irrelevant in the vacuum limit ($\sigma\rightarrow\infty$), since the radio wave in vacuum propagates through the background magnetic field with no interaction and no compression effect. A complete picture is revealed by the full time-dependent solution (section~\ref{relaxation}). It shows that the MHD flow pattern in the propagating wave packet is gradually attracted to the steady state, and this relaxation takes a significant time $\trel$ proportional to $\sigma$. For propagation times $t<\trel$, the steady state with large $C_\star$ is not reached; we evaluate the compression factor $C\ll C_\star$ in this situation.

These results allow one to investigate the FRB interaction with a magnetar wind in a broad range of radii. We describe the wind in section~\ref{wind}, and then investigate the compression front induced by the FRB in the wind. We identify two zones, with regular and stochastic heating, and find the transition radius $\rstoch$ between them. Compression fronts from FRBs emitted in each zone are described in sections~\ref{regular} and \ref{stochastic}. FRB damping by its interaction with the wind is investigated in section~\ref{damping}. Our conclusions are summarized in section~\ref{discussion}.


\section{Problem formulation}
\label{formulation}

\subsection{The role of magnetic field}

For an unmagnetized plasma, particle acceleration in a strong electromagnetic wave is a textbook problem \citep{Landau75} well known in the context of laser experiments \citep{Bulanov15}. It was also discussed in the context of early pulsar models \citep{Gunn71}. The wave accelerates initially static particles to the Lorentz factor $\gamma=1+a^2/2$. The particle oscillating in the wave has a large mean velocity aligned with the wave propagation direction. This mean velocity gives the plasma bulk Lorentz factor $\Gamma\approx a\gg 1$.

The problem qualitatively changes when the background plasma is magnetized \citepalias{Beloborodov22}. Now, in addition to the oscillating electromagnetic field of the wave, there is a background magnetic field $\bBbg$, which arrests the particle motion across $\bBbg$ and changes the frame where the particles have zero average momentum (the fluid frame). In the media surrounding FRB sources, $\Bbg$ is well below the radio wave amplitude $E_0$ at radii $r>\RLC$ considered in this paper. At the same time, $\Bbg$ is sufficiently strong to ``magnetize'' the plasma: the particle Larmor period is far smaller than the duration of the radio wave packet $T\sim 1$\,ms. Larmor rotation couples the plasma to $\Bbg$, and together they form an MHD fluid, which becomes heated and pushed by the radio wave.

Besides the wave strength parameter $a$, the problem of FRB-plasma interaction has two more dimensionless parameters: the background plasma magnetization $\sigma$ and the ratio of gyrofrequency $\omB=e\Bbg/\me c$ to the wave frequency $\om$. The parameters $a$ and $\omB/\om$ determine whether the plasma particles execute regular oscillations or experience stochastic heating in the radio wave. In addition to heating, the radio wave also deposits momentum, pushing the plasma along the wave direction. This effect depends on $\sigma$: $\sigma\gg 1$ implies that the effective bulk inertia of MHD fluid is significantly increased by its magnetic field, which can slow down its bulk acceleration. However, the acceleration effect may still be strong in waves with $a\gg 1$. 

Dynamics of a magnetized plasma exposed to the radio wave is easiest to calculate in the plane geometry, assuming a uniform magnetized background ahead of the wave packet. Details of the radio wave are not important; however, when simulating the plasma dynamics numerically, one needs to choose a concrete shape of the wave packet.

\subsection{Radio wave packet}

The wave speed is weakly affected by the presence of a tenuous plasma, and it will be taken equal to $c$ (this approximation is discussed in Appendix~\ref{vgr}). It is convenient to use the coordinate 
\beq
  \xi\equiv t-\frac{z}{c},
\eeq
where the $z$-axis is along the wave propagation direction. The radio wave packet is static in the $\xi$-coordinate and occupies a finite region $0<\xi<T$. As an example, consider a nearly harmonic plane electromagnetic wave with amplitude $E_0$, frequency $\om$, and duration $\tw\gg 2\pi/\om$. The wave electric field, 
\beq
\label{eq:Ew}
  E_{\rm wave}=E_0(\xi)\sin(\omega\xi),
\eeq 
oscillates with a high $\omega$ and a gradually varying amplitude $E_0(\xi)$; $E_0=0$ at the leading edge of the packet $\xi=0$, rises on a timescale $\xirise\gg \omega^{-1}$, and then drops back to zero at $\xi=T$.

Using $cE_{\rm wave}=-dA_{\rm wave}/d\xi$ one finds the wave potential at $\xi\omega\gg 1$: $A_{\rm wave}\approx A_0\cos(\omega\xi)$ where $A_0=c E_0/\omega$. So,
\beq
\label{eq:aw}
  a_{\rm wave}=a_0(\xi)\cos(\omega\xi), \qquad a_0=\frac{eE_0}{\me c\om},
\eeq
where $a_{\rm wave}\equiv eA_{\rm wave}/\me c^2$. The wave strength parameter is
\beq
\label{eq:a1}
     a\equiv \langle a_{\rm wave}^2\rangle^{1/2}=\frac{a_0}{\sqrt{2}}.
\eeq
 In numerical examples we will use packets with the envelope
\beq
\label{eq:packet}
 a(\xi)=a_{\rm max}\sin^2(\pi\xi/T), \qquad 0<\xi<T.
\eeq

The cold unperturbed medium ahead of the packet (at $\xi<0$) will be called ``upstream.'' The upstream medium is described by a uniform mass density $\rhou$ and a uniform magnetic field $\bB_{\rm u}$. We will focus on the case with $\bBu$ oriented perpendicular to the wave propagation direction (as expected in winds, which carry magnetic fields transverse to the radial direction).\footnote{The problem formulation with transverse $\bBbg$ is not special or degenerate in any way. One could also consider $\bBbg$ with arbitrary angle $\alpha$ with respect to the wave propagation direction. The general case with any $\alpha\neq 0$ is reduced to the propagation problem with $\alpha=\pi/2$ by using a new reference frame that moves along $\bBbg$ with Lorentz factor $\gamma=(\sin\alpha)^{-1}$ \citep{Beloborodov24}.}

Two possible polarizations of the radio wave can be considered: the X-mode ($\bEw\perp\bBu$) and the O-mode ($\bEw\parallel\bBu$). We will use below the results of \citetalias{Beloborodov22} for test-particle motion in the X-mode waves; similar results within numerical coefficients ${\cal O}(1)$ were recently published for the O-mode \citep{Huang24}. The X-mode has the oscillating magnetic field $\bBw$ parallel to the background field $\bBu$.

\subsection{Plasma drift through the wave packet}

The plasma behaves as an MHD fluid on timescales longer than the Larmor period. The fluid motion in the plane geometry with $\bBu$ perpendicular to the $z$-axis is fully described by the drift velocity parallel to the $z$-axis, 
\beq
  \bb_{\rm D}=\frac{\bvD}{c}=\frac{\overline\bE\times\overline\bB}{{\overline B}^2}. 
\eeq
Here $\overline\bE$ and $\overline\bB$ are the MHD fields; the bar signifies averaging over a timescale longer than the Larmor time (and longer than the wave oscillation period). The average fields can also be denoted as $\bEbg$ and $\bBbg$ because they serve as the effective background fields in the problem of test particle motion in the wave, as described below.

A natural choice for a reference frame in the wave-plasma interaction problem is such that the upstream plasma is at rest. When discussing wave propagation in a wind, we will also view the problem in the lab frame where the upstream plasma moves with a large speed $\betau>0$ along the wave propagation direction. In addition, frames with opposite motion, where the upstream plasma has $\betau<0$, may be useful in numerical simulations. In all these frames, $\bvD$ is parallel to the $z$-axis, both outside and inside the wave packet. The fluid four-velocity $\uD^\mu$ then has two non-zero components: 
\beq
   \uD^t=\gD=(1-\bD^2)^{-1/2}, \qquad \uD^z=\gD\bD.
\eeq
Dynamical equations for $\uD^\mu$ and final results will be formulated in a Lorentz-invariant way, so they will hold for any frame boosted along $z$ or, equivalently, any $\betau$.

One can also describe fluid motion using the quantity
\beq
\label{eq:kappa}
   \c\equiv \gD(1+\bD)=\frac{1}{\gD(1-\bD)}.
\eeq
This simplifies transformations between different frames moving along $z$. Relativistic addition of velocities corresponds to multiplication of $\c$, and changing $\bD\rightarrow -\bD$ corresponds to $\c\rightarrow \c^{-1}$. Doppler transformation of the wave frequency satisfies $\omega/\c={\rm inv}$. Transformation of other quantities follow from $a={\rm inv}$ (strength parameter of the wave) and $\omega\xi={\rm inv}$ (phase of the wave). In particular, transformations of the wave electric field $E_{\rm wave}$, wave energy density $\Uw$, and coordinate $\xi$ measured in two reference frames $\KF$ and $\tKF$ are
\beq
\label{eq:transform}
      \frac{\tilde{E}_{\rm wave}}{\Ew}=\frac{\xi}{\tilde\xi}=\frac{\tom}{\om}=\frac{\tilde\c}{\c}, 
      \qquad \frac{\tUw}{\Uw}=\frac{\tilde{\c}^2}{\c^2}.
\eeq 
Hereafter, we choose $\tKF$ as the local rest frame of the fluid,\footnote{Throughout this paper, we use tilde for quantities measured in the fluid frame $\tKF$. Tilde will be omitted if this does not lead to confusion: we omit tilde for thermodynamic quantities (e.g. pressure $P$ and enthalpy $w$, which are defined only in frame $\tKF$).}
which corresponds to
\beq 
  \tilde{\beta}_{\rm D}=0, \qquad \tilde\c=1.
\eeq

Plasma with any speed $\vD=dz/dt$ (in any frame) changes its $\xi$-coordinate with rate 
\beq
  \frac{d\xi}{dt}=1-\bD>0.
\eeq
Thus, MHD fluid always moves toward increasing $\xi$ through the wave packet. When viewed in the rest frame of the upstream medium, the initially static fluid enters the wave packet at $\xi=0$ and develops speed $\bD\neq 0$ at $\xi>0$. We wish to find the fluid motion in the packet, $\bD(\xi)$, and the fluid compression factor $C(\xi)=\rho(\xi)/\rhou$.


\section{Plasma heating by the radio wave}
\label{heating}

The response of a magnetized plasma to a propagating radio wave packet can be disentangled if we first focus on the ``micro level'' and consider the motion of individual particles in the local fluid frame $\tKF$. In this frame, the fluid is at rest and the wave-plasma interaction may be viewed as a local heating effect: the strong high-frequency wave energizes particles to Lorentz factors $\tg\gg 1$. This effect was calculated in \citetalias{Beloborodov22} by tracking the motion of test particles. The heating depends on the wave strength parameter $a$. For smooth packets, such that $a$ varies on a scale $\Delta\tilde{\xi}\gg a\tomB$ (typically satisfied by FRBs), the results for $\tg$ may be summarized as follows (see also Appendix~\ref{app:motion}).

(1) Waves with frequency $\tom$ exceeding the gyrofrequency $\tomB=e\tBbg/\me c$ drive regular particle oscillations with the average Lorentz factor
\beq
\label{eq:tg1}
    \langle\tg\rangle=\sqrt{1+a^2}.
\eeq
The local $\langle\tg\rangle$ in the packet simply tracks the local strength parameter of the wave $a$. 

(2) In the other regime, $\tomB\simgt \tom$, the wave was found to drive stochastic particle acceleration. Then, ensemble average $\langle\tg\rangle_{\rm ens}$ for particles inside a fluid element quickly exceeds $a$ and keeps growing as the fluid moves through the wave packet (until radiative losses limit $\tg$ or the fluid exits behind the packet). The growth of $\langle\tg\rangle_{\rm ens}$ is described by the following equation (\citetalias{Beloborodov22},\citealt{Beloborodov24})
\beq
\label{eq:stochastic}
   \frac{d\langle \tg\rangle_{\rm ens}}{d\tilde t} 
   = \chi \frac{a^2\tomB^{1/3}\tom^{2/3}}{\langle\tg\rangle_{\rm ens}^{4/3}},
\eeq
where $\chi\sim 1$ is a numerical coefficient, and all quantities are measured in the fluid rest frame $\tKF$. Energy deposition by the radio wave via stochastic heating of particles may be viewed as a process of synchrotron absorption. The mean energy gain per Larmor rotation by a particle with $\tg\gg a$ was calculated by \cite{Lyubarsky18}, which may be used to guess the scaling in \Eq~(\ref{eq:stochastic}). Calculation of $d\langle\tg\rangle_{\rm ens}/dt$ must however track the evolving particle distribution function $f(\tg)$, shaped by stochastic heating (Appendix~\ref{stochastic_orbit}). The final result is described by \Eq~(\ref{eq:stochastic}) with $\chi\approx 0.8$. Below we drop the angular brackets in $\langle\tg\rangle$ for simpler notation, so $\tg$ will denote the average particle Lorentz factor in a fluid element.

Calculations in \citetalias{Beloborodov22} were performed assuming that the background magnetic field $\bBbg$ is static, i.e. there is no background electric field $\bEbg$ and hence no average particle drift: $\bvD/c=\bEbg\times\bBbg/\Bbg^2=0$. This corresponds to using the frame $\tKF$ where the magnetized fluid is at rest --- the energized particles have no net momentum as they gyrate in a static magnetic field. Note that the wave fields $\bEw$ and $\bBw$ oscillate around zero and do not contribute to the MHD fields $\overline{\bE}$ and $\overline{\bB}$ (averaged over a timescale longer than the particle Larmor period and the wave period). So, the MHD fields serve as the background fields in the test-particle calculations:
\beq
   \overline\bE=\overline{\bEbg +\bEw} = \bEbg, \quad
   \overline\bB=\overline{\bBbg +\bBw} = \bBbg.
\eeq 

In general, $\overline\bE\neq 0$ which corresponds to a moving MHD fluid $\vD\neq 0$. If $\vD$ varies (the plasma is accelerated in the wave packet), no global inertial frame exists where the fluid would be static, and one has to consider the heating effect of the radio wave in a moving MHD fluid. Then, the test particle motion should be examined with $\Ebg\neq 0$, as done in this section. 

One goal of our calculations below is to generalize the threshold condition for stochastic heating $b\equiv \tomB/\tom\gtrsim 1$ so that it applies to accelerating fluids. If the upstream plasma has $\bup\gtrsim 1$, stochastic heating is triggered immediately in the leading part of the wave packet once $a$ exceeds $\sim 1$ \citepalias{Beloborodov22}. If the upstream plasma has $\bup\ll 1$, stochastic heating can still occur, but deeper inside the wave packet where $b$ is sufficiently increased by fluid acceleration and compression. This can occur after the wave strength parameter has reached a high $a\gg 1$, which changes the threshold for stochastic heating.

Remarkably, in an accelerating fluid, \Eqs~(\ref{eq:tg1}) and (\ref{eq:stochastic}) still hold, i.e. the evolution of $\tg$ in both regimes remains unchanged. This fact will be illustrated below by direct calculations of particle motion in a wave packet with a non-static background field, which defines the fluid motion. We will prescribe smoothly varying fields $\bBbg(\xi)$ and $\bEbg(\xi)$ (varying on a scale much greater than the particle Larmor radius) and repeat the calculation of \citetalias{Beloborodov22} for test particles. 

The background fields will be parameterized as follows
\beq
\label{eq:EB_bg}
   \Bbg=\frac{\c^2+1}{\cu^2+1}\Bbg^{\rm u}, \qquad \Ebg=\frac{\c^2-1}{\cu^2+1}\Bbg^{\rm u}.
\eeq 
In the simulations, we will use the frame where the upstream plasma is at rest; then, $\Ebg^{\rm u}=0$ and $\cu=1$. Fluid acceleration inside the wave packet is specified by a chosen function $\c(\xi)$. For the present purposes, \Eq~(\ref{eq:EB_bg}) may be viewed as an arbitrary prescription to test heating of fluid with non-uniform motion $\bD(\xi)$, which is related to $\c(\xi)$ by \Eq~(\ref{eq:kappa}). As explained in next sections, any steady pattern of fluid motion in the wave packet actually satisfies \Eq~(\ref{eq:EB_bg}). It also satisfies $\trho/\c=const$ (uniform across the packet), where $\trho=\rho/\gD$ is the fluid proper density. Thus, the adopted prescription corresponds to a steady flow pattern with the proper compression factor $\trho/\trhou=\c/\cu$. The magnetic field $\tBbg$ is frozen in the fluid and compressed by the same factor $\c/\cu$ while $\Bbg=\gD\tBbg$ obeys \Eq~(\ref{eq:EB_bg}).

Our simulations will track test particles initially located in the static, cold upstream region exposed to the propagating wave packet. The particle motion in the wave packet is found by solving the dynamical equation,
\beq
\label{eq:dyn}
    mc\,\frac{d\boldsymbol{u}}{dt}=e\left(\bE + \bb\times\bB\right), 
\eeq
where vector $\boldsymbol{u}=\gamma\bb$ is the three-dimensional spatial part of the particle four-velocity. Radiative losses are neglected in \Eq~(\ref{eq:dyn}). The fields $\bE(\xi)$ and $\bB(\xi)$ include the background and wave fields: $\bE=\bEbg+\bEw$ and $\bB=\bBbg+\bBw$.\footnote{Notation here differs from \citetalias{Beloborodov22} where $\bE$ and  $\bB$ denoted the wave fields and the dynamical equation assumed $\bEbg=0$.}

\subsection{Regular oscillations}
\label{regular_heating}

Let us prescribe $\Bbg(\xi)$ and $\Ebg(\xi)$ by choosing $\c^2$ of the form
\beq
\label{eq:bg}
   \frac{\c^2}{\cu^2}=1+ \zeta a^2,
\eeq
where $\zeta$ is a constant coefficient. The simulation will be done in the reference frame where the upstream plasma is at rest ($\cu=1$). The case of $\zeta=0$ corresponds to a static magnetic background throughout the packet ($\c=1$, $\Ebg=0$); in this case, the particle motion is similar to that found in \citetalias{Beloborodov22}. When $\zeta\neq 0$, the increasing $\c>1$ means $\bD>0$ and the appearance of additional electric force $e\Ebg$, which changes the particle motion. The prescribed $\c(\xi)$ gives the fastest drift of the fluid in the middle of the packet where $a(\xi)$ peaks; the fluid is static ahead and behind the wave packet where $a=0$. 

We have calculated three sample models with $\zeta=0$, 0.4, and 1. In all three models, the radio wave packet has the amplitude $a_{\max}=4$, and the upstream plasma has the gyro-frequency $\tomBu=0.03\om$, where $\omega$ is the wave frequency measured in the upstream rest frame. These parameters represent an FRB propagating through the magnetar wind at radii $r\sim 3\times 10^{12}$\,cm, as discussed in section~\ref{wind} below. Another parameter is the number of wave oscillations in the packet, $N=\omega T/2\pi$. Real FRBs have $N\sim 10^6,$ and the presented simulations have $N=10^3$. We have also run models with different $N$ and verified that the results are independent of $N$ as long as $N\gg a\tom/\tomB$ (this is simply the condition that the packet is much wider than the particle Larmor radius).

\begin{figure}[t]
\includegraphics[width=0.47\textwidth]{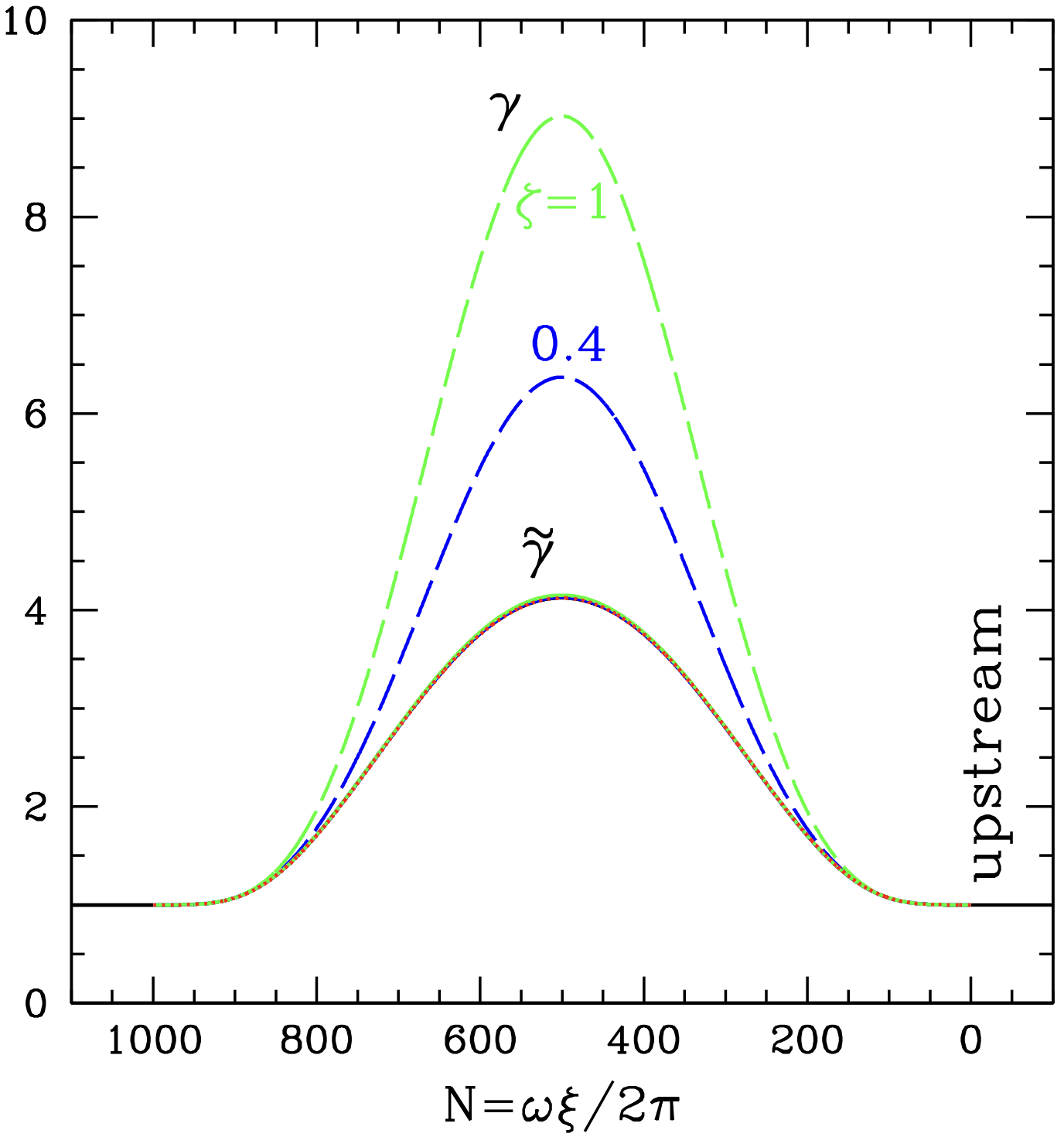}
\caption{Lorentz factor  $\gamma$ of a test particle moving through the wave packet (averaged over the wave oscillation period). The upstream plasma is cold ($\tg=1$) and static ($\cu=1$); it has gyro-frequency $\tomBu=0.03\omega$ measured in the fluid frame. The test particle moves through the wave packet in the coordinate $\xi=t-z/c$ from right to left.  The simulated packet consists of $10^3$ oscillations with an envelope described by \Eq~(\ref{eq:packet}) with $a_{\max}=4$, and the background field is described by \Eqs~(\ref{eq:EB_bg}) and (\ref{eq:bg}). Different colors correspond to models with $\zeta=0$ (static background, $\Ebg=0$; black), $\zeta=0.4$ (blue), and $\zeta=1$ (green). For each model, dashed curve shows $\gamma$ measured in the fixed frame (where the upstream plasma is at rest), and solid curve shows $\tg$ measured in the local fluid rest frame. All three solid curves overlap, demonstrating that $\tg(\xi)$ is independent of the local fluid velocity $\bD=\Ebg/\Bbg$ and follows the solution $\tg^2=1+a^2(\xi)$ (red dotted curve, overlaping the solid curves).}
\label{fig:test}
\end{figure}

For each model, we have solved for the particle motion in the radio wave packet, and then calculated its Lorentz factor averaged over the wave oscillation period: $\gamma$ in the lab frame and $\tg$ in the local fluid rest frame. Figure~\ref{fig:test} shows the average $\gamma$ and $\tg$ as functions of position $\xi$ in the packet. It demonstrates that $\tg=\sqrt{1+a^2}$ holds in all three models ($\zeta=0$, 0.4, 1). In the fluid frame, the particle performs  same regular oscillations, independent of the fluid bulk acceleration in the packet. Since fluid motion $\c$ varies on scales much larger than the Larmor scale, Larmor rotation is not excited \citepalias{Beloborodov22}. The particle orbit measured in the local fluid frame is nearly identical to the canonical ``8''-shaped orbit, same as found in the absence of any background field (Appendix~\ref{app:motion}). 

\subsection{Stochastic heating}
\label{stochastic_heating}

The onset of stochastic heating is controlled by two parameters: $a$ and 
\beq
  b\equiv \frac{\tomB}{\tom}.
\eeq
As an example, consider a radio wave packet with amplitude $a_{\max}=30$ propagating in an ambient plasma with gyrofrequency $\tomBu=0.2\tomu$, so the upstream plasma has $b_{\rm u}=0.2$. These parameters may represent an FRB propagating through a magnetar wind at $r\sim 5\times 10^{11}$\,cm (section~\ref{wind}). In this case, the fluid is accelerated and compressed inside the wave packet by a large factor $\c/\cu\gg 1$, and $b\gg 1$ can be reached inside the packet. Indeed, \Eq~(\ref{eq:EB_bg}) implies compression of $\tBbg$ by the factor of $\c/\cu$, so the gyrofrequency measured in the local fluid frame $\tKF$ is increased to $\tomB=(\c/\cu)\tomBu$. At the same time, the wave frequency measured in the fluid frame is reduced from $\tomu=\omega/\cu$ to $\tom=\om/\c$. This gives $b=(\c/\cu)^2b_{\rm u}$. 

The threshold condition for the onset of stochastic heating can be stated in the form $b>\bs(a)$, as discussed in section~\ref{transition} below. The transition to stochastic heating occurs for any profile of $\c(\xi)$ as long as it reaches $b>\bs$, i.e.  $\c/\cu$  exceeds $(\bs/b_{\rm u})^{1/2}$. In our example simulations, we prescribe $\c(\xi)$ using a modified version of \Eq~(\ref{eq:bg}):\footnote{We use the simplest modification of \Eq~(\ref{eq:bg}) that satisfies $d\c/d\xi\geq 0$. This condition is motivated by the results of later sections that describe self-consistent calculations of $\c(\xi)$. Unlike regular oscillations (where $\tg$ tracks the rise and drop of $a(\xi)$ in the wave packet), stochastic heating gives a monotonically growing expectation value $\tg(\xi)$: $d\tg/d\xi\geq 0$, leading to $d\c/d\xi\geq 0$.}
\begin{eqnarray}
\label{eq:bg1}
  \frac{\c^2}{\cu^2}=\left\{\begin{array}{cc}
\vspace*{1mm}
  1+\zeta a^2 & \qquad \xi<T/2 \\
  1+\zeta a_{\max}^2 & \qquad \xi>T/2
              \end{array}\right.
\end{eqnarray}
The results will be shown for two sample models: $\zeta=0$ (no fluid acceleration in the packet) and $\zeta=1$. 

\begin{figure}[t]
\includegraphics[width=0.47\textwidth]{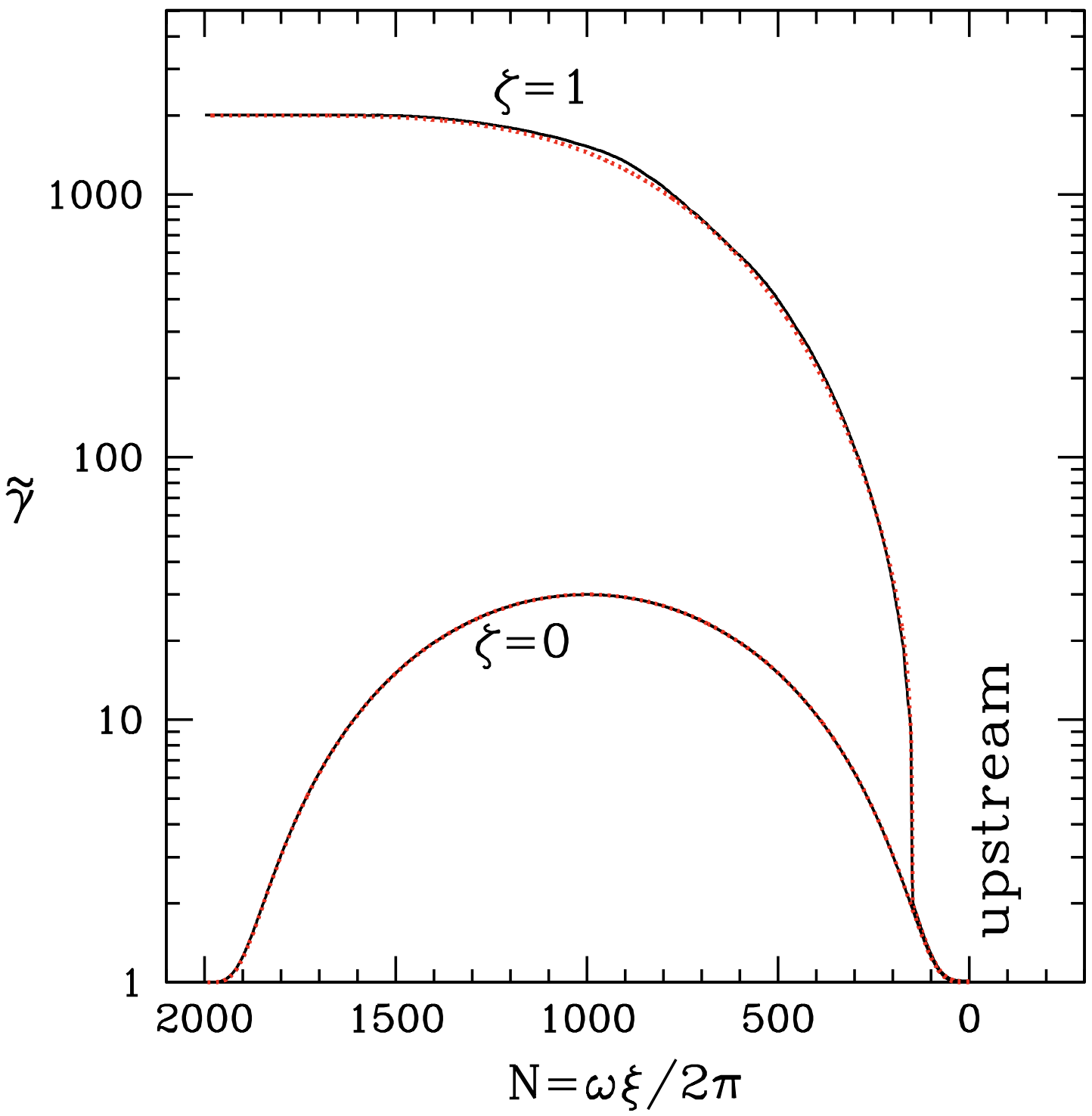}
\caption{Average Lorentz factor $\tg$ (measured in the local fluid rest frame) of particles interacting with the wave packet with amplitude $a_{\max}=30$. The upstream plasma is at rest ($\cu=1$) and has a small temperature $kT=10^{-2}\me c^2$; its gyro-frequency is $\tomBu=0.2\tomu$. The average $\tg$ is calculated by tracking $4\times 10^3$ test particles. Background fields $\Bbg$ and $\Ebg$ are described by \Eqs~(\ref{eq:EB_bg}) and (\ref{eq:bg1}). The model with $\zeta=0$ corresponds to no fluid acceleration in the wave packet and shows regular oscillations with $\tg=\sqrt{1+a^2}$. The model with $\zeta=1$ demonstrates the transition to stochastic heating caused by the strong acceleration and compression of the fluid. Red dotted curves show the analytical predictions for regular oscillations ($\tg=\sqrt{1+a^2}$) and stochastic heating (\Eq~(\ref{eq:tg_stochastic}) with $\chi=0.8$).}
\label{fig:test_stoch}
\end{figure}

Measurements of stochastic heating require calculations of many realizations of particle motion in the wave packet. We draw 4000 upstream particles from an initial Maxwellian distribution with a small temperature $kT=10^{-2}\me c^2$, seeding initial perturbations for the stochastic behavior. The particle motion is followed for each of the 4000 realizations, and the result is averaged to find the mean value $\tg(\xi)$. The results are shown in Figure~\ref{fig:test_stoch}.

In the model with $\zeta=0$, the ratio $b\equiv\tomB/\tom=0.2$ stays uniform and small across the wave packet. Then, the stochastic heating does not occur: the plasma everywhere performs regular oscillations with $\tg=\sqrt{1+a^2}$. By contrast, in the model with $\zeta=1$, the background field is subject to strong compression, $\c/\cu=\sqrt{1+a^2}$, which changes the particle response to the radio wave. In the leading part of the packet, where $\c/\cu$ is still small, the particles perform regular oscillations  with $\tg=\sqrt{1+a^2}$, same as in the model with $\zeta=0$. The transition to stochastic heating is observed at $\xi_{\rm stoch}\approx 140\times 2\pi/\omega$, where $\tg(\xi)$ shows a steep growth. 

This steep growth of $\tg$ closely follows the analytical expectation for stochastic heating. It is stated in \Eq~(\ref{eq:stochastic}), which describes $\tg(\tilde t)$, and one can easily find the corresponding $\tg(\xi)$. The fluid proper time $\tilde{t}$ is related to time $t$ in the lab frame by $d\tilde{t}=dt/\gD$. During time $dt$, the fluid element moves in $\xi$ by $d\xi=(1-\bD)dt$, so
\beq
   d\tilde{t}=\frac{d\xi}{\gD(1-\bD)}=\c\, d\xi.
\eeq
Using also $\tom=\om/\c$ and $\tomB=\tomBu\c/\cu$, we rewrite \Eq~(\ref{eq:stochastic}) as
\beq
\label{eq:tg_stochastic}
     \tg^{4/3}\,\frac{d\tg}{d\xi} = \chi\, a^2 \c^{2/3} \left(\frac{\tomBu\om^2}{\cu}\right)^{1/3}.
\eeq
For any chosen $\c(\xi)$, \Eq~(\ref{eq:tg_stochastic}) can be integrated for $\tg(\xi)$. As one can see in Figure~\ref{fig:test_stoch}, this analytical calculation reproduces the mean value $\tg(\xi)$ found in the full dynamical simulation with the ensemble of 4000 particles. We have also calculated models with different $a_{\max}$, $\tomBu/\tomu$, and $\zeta$, and found that \Eq~(\ref{eq:tg_stochastic}) with $\chi\approx 0.8$ well describes stochastic heating in all of them.

\begin{figure}[t]
\includegraphics[width=0.47\textwidth]{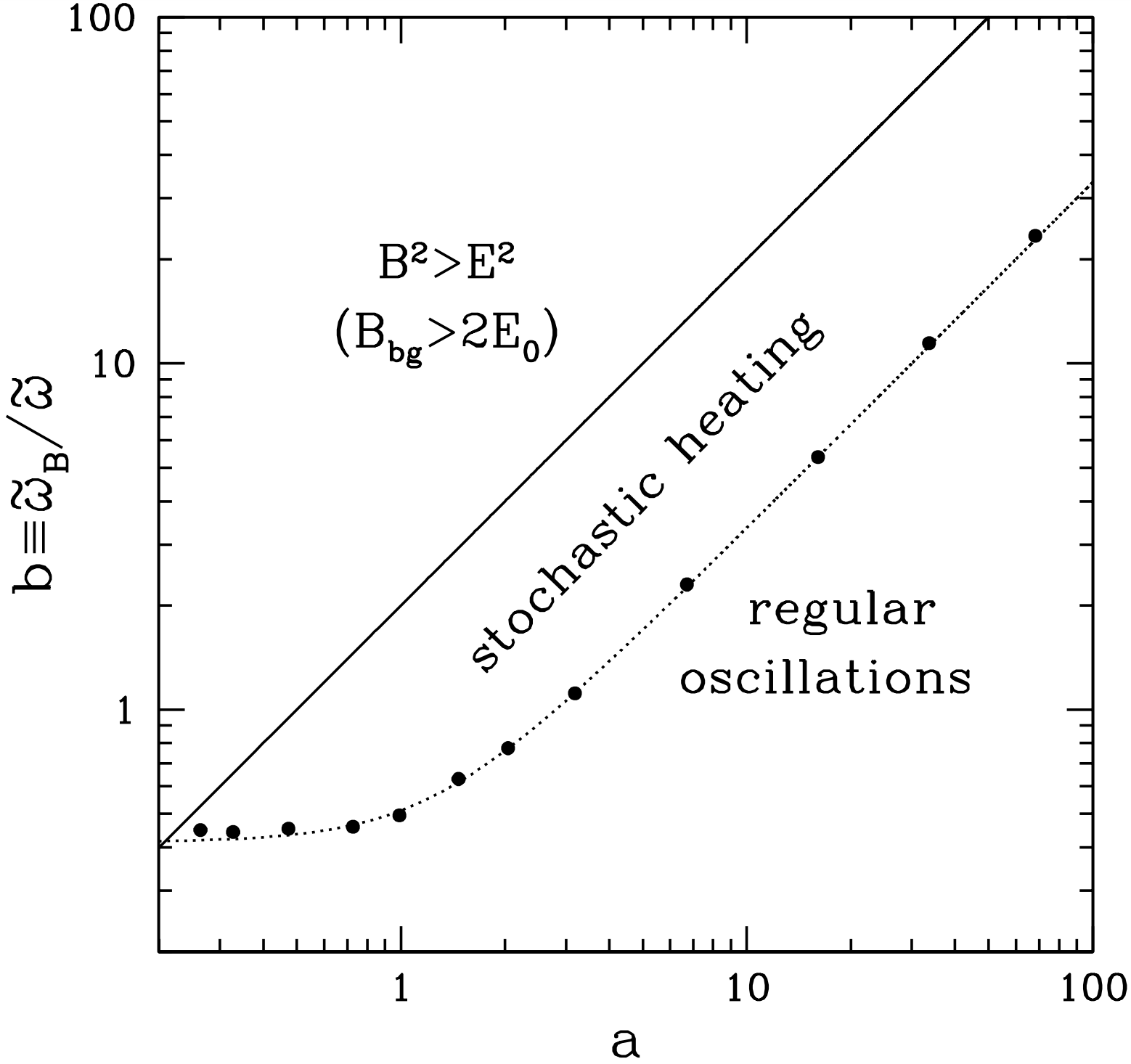}
\caption{Transition curve separating the regimes of regular oscillations and stochastic heating on the $a$-$b$ plane. Black circles show the results of 12 simulations: each circle shows $(a,b)$ at which the transition from regular oscillations to stochastic heating occurred as the particles moved through the wave packet. Dotted curve shows $b=(1/3)\sqrt{1+a^2}$. The region above the solid curve satisfies $b>2\sqrt{2}\,a$, which corresponds to wave amplitude $\tE_0<\tBbg/2$. In this region, the electromagnetic field satisfies $E^2<B^2$ throughout the radio wave oscillation, and stochastic heating is not expected. The radio wave cannot compress the background far above the solid line, as this would require energy exceeding the available energy of the wave.}
\label{fig:trans}
\end{figure}

As illustrated in Figures~\ref{fig:test} and \ref{fig:test_stoch}, the particle response to the radio wave admits a simple analytical description in frame $\tKF$ regardless of how this frame moves in the radio wave packet. This universality is foreseeable. The speed of frame $\tKF$ varies on timescales much longer than the particle Larmor period; so, in the problem of particle response to the radio wave, $\tKF$ is not too different from an inertial frame. As a result, $\tg$ evolves with the same rate as in a non-accelerating fluid: it follows \Eq~(\ref{eq:tg1}) in the regular oscillation regime and \Eq~(\ref{eq:tg_stochastic}) in the stochastic heating regime.

The large Lorentz factor $\tg\gg 1$ may be thought of as a heating effect, giving the fluid a relativistic effective enthalpy and increasing its effective inertial mass. Note that the Larmor frequency $\tomL=\tomB/\tg$ remains below the wave frequency $\tom$ for both regular oscillation and stochastic heating regimes \citepalias{Beloborodov22}. The ratio $\tomL/\tom$ reaches a maximum value of $\approx 1/3$ at the transition between the two regimes, which is described below.

\subsection{Transition curve on the $a$-$b$ plane} 
\label{transition}

Each test-particle simulation that shows the transition to stochastic heating inside the wave packet at some point $\xi_{\rm stoch}$ gives a measurement of $a(\xi_{\rm stoch})$ and $b(\xi_{\rm stoch})$. These measurements are shown in Figure~\ref{fig:trans} for a set of simulations with various $b_{\rm u}$, $a_{\max}$, and $\zeta$. We kept $b_{\rm u}$ below 0.4 in all cases; then, stochastic heating is not activated without accelerating/compressing the background. Compression enables the transition by increasing $b$ from $b_{\rm u}$ by the factor $b/b_{\rm u}=\c^2/\cu^2=1+\zeta a^2$.

As one can see in Figure~\ref{fig:trans}, the transition occurs on a well defined curve in the $a$-$b$ plane. It is independent of the shape of the wave packet $a(\xi)$ or the profile of background compression in the packet (here parameterized by $\zeta$);  it is also independent of $b_{\rm u}$. Basically, the transition from regular oscillations to stochastic heating is controlled by local $a$ and $b$. This is expected, since the local plasma response to the wave viewed in the fluid frame is independent of the drift of this frame or the global shape of the wave packet.

The transition curve extends from $a\sim 0.3$ to $a\gg 1$, and approximately follows the relation 
\beq
\label{eq:trans}
  b=\bs\approx \frac{1}{3} \sqrt{1+a^2} \qquad {\rm (transition)}.
\eeq
At $b<\bs$, the plasma performs regular oscillations in the radio wave with frequency $\tom$ while  Larmor rotation (with frequency $\tomL=\tomB/\tg=\tomB/\sqrt{1+a^2}<\tom/3$) is not excited. Excitation of Larmor rotation and transition to stochastic heating occur when $\tomL$ exceeds $\tom/3$ (Larmor radius $\tilde{r}_{\rm L}=c/\tomL$ exceeds half wavelength $\pi c/\tom$), which corresponds to $b>\bs$. At this point, the quick growth of $\tg$ is triggered, with $\tg$ reaching very high values and $\tomL$ decreasing far below $\tom$.

Within a numerical factor, \Eq~(\ref{eq:trans}) can be obtained by combining $\tg\sim a$ (achieved in regular oscillations) with the condition for triggering stochastic heating $\tg\gtrsim \tg_\star\approx(a^3/b)^{1/2}$ \citepalias{Beloborodov22}. The transition to stochastic heating at $\tg\sim \tg_\star$ may be thought of as the onset of synchrotron absorption of the wave.


\section{MHD equations for fluid motion}
\label{MHD}

In the test-particle numerical experiments described above, we prescribed an arbitrary pattern of the fluid drift speed $\bD(\xi)$ or, equivalently, $\c(\xi)$. Our goal now is to find the actual pattern of fluid motion in the radio wave packet. It can be quantified by the profiles of two quantities: $\kappa$ (related to $\bD$) and $\tC=\trho/\trhou$ (the proper compression factor). They turn out equal, $\tC=\kappa$, when a steady pattern of fluid motion is established in the propagating wave packet, i.e. when $\kappa$ and $\trho$ are functions of $\xi$ only. The full problem, however, involves evolving profiles. Therefore, we will formulate general MHD equations for $\kappa(t,\xi)$ and $\trho(t,\xi)$, allowing time dependence. We first state the equations in the usual variables $t,z$ and later switch to the more convenient coordinates $t,\xi$.

\subsection{Conservation laws}
\label{cons}

Fluid motion in the wave packet obeys conservation of mass, energy, and momentum. They are formulated using  mass four-flux $\trho \uD^\nu$ and stress-energy tensor $T^{\mu\nu}$. Only the $t$ and $z$ components are relevant in the plane-parallel problem.

Mass conservation for an MHD flow (assuming negligible $e^\pm$ creation/annihilation) reads
\beq
\label{eq:mass}
   \partial_\mu(\trho \uD^\mu)=\partial_t\rho+\partial_z(\rho\vD)=0.
\eeq
Contributions to $T^{\mu\nu}$ are made by the plasma, the MHD electromagnetic field, and the radio wave:
\beq
\label{eq:Uw}
   T^{\mu\nu}=T^{\mu\nu}_{\rm p}+T^{\mu\nu}_{\rm f}+T^{\mu\nu}_{\rm w}.
\eeq
The relevant components of $T^{\mu\nu}$ are $T^{tt}$, $T^{tz}$, and $T^{zz}$. In particular, the stress-energy tensor of the radio wave is 
\beq
  T^{zz}_{\rm w}=T^{tz}_{\rm w}=T^{tt}_{\rm w}=\Uw=\frac{\langle E_{\rm wave}^2\rangle}{4\pi}=\frac{\me\, \om^2 a^2}{4\pi r_e},
\eeq
where $r_e=e^2/\me c^2$ is the classical electron radius.

Consider now the relevant components of the stress-energy tensor of the MHD electromagnetic field:
\beq
  T^{tt}_{\rm f}=T^{zz}_{\rm f}=\frac{B^2+E^2}{8\pi},  \qquad  T^{tz}_{\rm f}=\frac{EB}{4\pi}.
\eeq
Hereafter, we drop the bar in $\overline{E}$ and $\overline{B}$ --- in all remaining sections, $\bE$ and $\bB$ will have the meaning of MHD fields averaged over the Larmor timescale. $T^{\mu\nu}_{\rm f}$ may be re-written in a form similar to ideal fluid: 
\beq
\label{eq:Tf}
   T^{\mu\nu}_{\rm f}=(\Uf+\Pf)\uD^\mu \uD^\nu+\eta^{\mu\nu}\Pf \qquad (\mu,\nu=t,z),
\eeq
where
\beq
   \Uf=\Pf=\frac{\tB^2}{8\pi}, \qquad \tB=\frac{B}{\gD}.
\eeq
Note that the MHD electric field vanishes in the fluid frame (there is no plasma drift in frame $\tKF$ by  definition). The magnetic field in the fluid frame $\tB$ controls the energy density and effective pressure $\Pf=\Uf$. Magnetic flux freezing in each fluid element implies
\beq
   \frac{\tB}{\Bu}=\frac{\trho}{\rhou}=\frac{\sigma}{\sigu},
   \qquad \sigma\equiv\frac{\tB^2}{4\pi\trho c^2}.
\eeq

The remaining component of the system is the plasma, which is energized by the radio wave. The plasma stress-energy tensor $\Tp^{\mu\nu}$ is derived in Appendix~\ref{app:T} for both possible regimes: stochastic heating and regular oscillation in the wave (\Eqs~(\ref{eq:Tstream_av}) and (\ref{eq:Tp}), respectively). The total MHD stress-energy tensor is  
\beq
   T^{\mu\nu}_{\rm MHD}=T^{\mu\nu}_{\rm p}+T^{\mu\nu}_{\rm f}.
\eeq

Interaction of the wave with the MHD fluid obeys energy and momentum conservation, 
\beq
\label{eq:cons}
  \partial_\nu\left(T^{\mu\nu}_{\rm MHD}+T^{\mu\nu}_{\rm w}\right)= - n  \dPe^\mu, 
\eeq
where $n=\rho/\me$ is the particle number density, and $\dPe^\mu=(\dEe/c,\dot{\boldsymbol{\P}}_e)$ is the average emission rate of four-momentum per particle, which is derived in Appendix~\ref{app:losses}. It is given by \Eqs~(\ref{eq:dE_dt_stream}), (\ref{eq:dP_dt_stream}) in the regular oscillation regime and by \Eqs~(\ref{eq:dE_dt_stochastic}), (\ref{eq:dP_dt_stochastic}) in the stochastic heating regime.

\subsection{MHD equations in coordinates $t,\xi$}

The conservation laws can also be stated using fluxes through the surfaces of $\xi=const$, which move with speed $c$. This will considerably simplify the analysis of wave-fluid interaction.

The fluid speed in the $\xi$-coordinate is $d\xi/dt=1-\bD$, so the mass flux through a surface of $\xi=const$ is 
\beq 
  F_{\rm m}=\rho(c-\vD)=\frac{c\trho}{\c}.
\eeq
Mass conservation in the $t,\xi$-coordinates reads
\beq
   \left.\frac{\partial \rho}{\partial t}\right|_{\xi}  + 
    \left.\frac{\partial F_{\rm m}}{\partial\xi}\right|_t = 0.
\eeq

Similarly, the flux of four-momentum $\P^{\mu}$ through a surface $\xi=const$ is $F^\mu=T^{\mu t}-T^{\mu z}$. Conservation of four-momentum (\Eq~\ref{eq:cons}) in the $t,\xi$-coordinates becomes\footnote{This equation can also be derived formally by transforming tensor $T^{\mu\nu}$ to new coordinates: $(t,z)\rightarrow(t,\xi)$. The resulting $T^{t\xi}$ and $T^{z\xi}$ are the energy and momentum fluxes crossing $\xi=const$.}
\beq
\label{eq:Tcons1}
    \left.\frac{\partial T^{\mu t}}{\partial t}\right|_{\xi} 
    + \frac{\partial}{\partial\xi}\left(T^{\mu t} - T^{\mu z} \right)_t  = - cn \dPe^\mu.
\eeq
Note that the wave is static in $\xi$, so it creates no flux of energy or momentum across a surface of $\xi=const$:\footnote{This statement holds for waves propagating with speed $c$. The exact propagation speed $\vgr<c$ (Appendix~\ref{vgr}) implies that the radio wave carries a non-zero energy flux in $\xi$: $F_{\rm w}=\Uw(c-\vgr)$. In particular, in the fluid frame $\tKF$, $\tilde F_{\rm w}\sim c\tUw \tom_{\rm p}^2/2\tom^2\sim \trho c^3 a^2/2\tg$. One may neglect $\tilde F_{\rm w}$ compared to flux $\tilde{F}$ created by the MHD fluid (effectively setting $\vgr=c$) if $a^2/\tg \ll h$, where $h$ is given in \Eq~(\ref{eq:h}) below. This condition turns out satisfied if the plasma has $\sigu\gg 1$.}
\beq
   F^\mu_{\rm w}=T^{\mu t}_{\rm w}-T^{\mu z}_{\rm w}=\Uw-\Uw=0  \qquad (\mu=t,z).
\eeq
Then, only the MHD fluid contributes to the fluxes.

\subsection{Conservation of $\Q^\pm$}
\label{Qpm}

Instead of conservation of energy $\E$ and momentum $\P$ it will be helpful to state conservation of quantities
\beq
   \Q^\pm=\E\pm c\P. 
\eeq
Conservation of $\Q^\pm$ is given by the sum/difference of equations stating energy and momentum conservation (\Eq~(\ref{eq:Tcons1}) with $\mu=t$ and $z$):
\begin{align}
\label{eq:Q+}
   \frac{\partial}{\partial t} (T^{tt} + T^{tz})
   + \frac{\partial}{\partial\xi} ( T^{tt} - T^{zz}) &= - \dot\Q^+, \\
\label{eq:Q-}
       \frac{\partial}{\partial t} (T^{tt} - T^{tz})
   +\frac{\partial}{\partial\xi} ( T^{tt} -2T^{tz} + T^{zz}) &= - \dot\Q^-,
\end{align}
where $\dot\Q^\pm=cn(\dPe^t\pm \dPe^z)$.

Note that the stress-energy tensor of the radio wave $T_{\rm w}^{\mu\nu}$ cancels out in the conservation law for $Q^-$ (\Eq~\ref{eq:Q-}), so all terms $T^{\mu\nu}$ in this equation can be replaced with $T^{\mu\nu}_{\rm MHD}$. Effectively, it states conservation of $Q^-$ carried by the MHD fluid. This fact has a simple physical reason: the radio wave with energy $\E$ and momentum $\P=\E/c$ carries $\Q^-=0$ and cannot exchange $\Delta Q^-\neq 0$ with the MHD fluid. Whenever the radio wave loses energy $\Delta \E$, it also loses momentum $\Delta\P=\Delta \E/c$, so
\beq
   \Delta \Q^- = \Delta \E - c\,\Delta\P = 0.
\eeq
This observation significantly simplifies the problem. As long as the propagating radio wave packet $a(\xi)$ is approximately steady (i.e. the wave is not damped), the fluid dynamics in the packet is fully described by two equations that express conservation of mass and $\Q^-$. The remaining independent equation for $Q^+$ is needed if one aims to explicitly follow the gradual evolution of the wave packet $a(t,\xi)$. It is not needed for FRBs that are weakly affected by their interaction with the wind. Here, we will assume this condition to be satisfied, and later (in section~\ref{damping}) will estimate the parameter space where strong FRB damping is expected.

Another observation concerns the form of the plasma stress-energy tensor $\Tp^{\mu\nu}$ derived in Appendix~\ref{app:T}. In the stochastic heating regime, it takes the usual ideal-fluid form (\Eq~\ref{eq:Tp}). By contrast, in the regular oscillation regime, $\Tp^{\mu\nu}$ looks different because of the first term in \Eq~(\ref{eq:Tstream_av}). However, this term is the same in all three components $\Tp^{tt}$, $\Tp^{tz}$, $\Tp^{zz}$ and drops out from \Eq~(\ref{eq:Q-}). Thus, effectively, only the second term in $\Tp^{\mu\nu}$ contributes to the $Q^-$ equation, i.e. the plasma stress-energy tensor can be replaced by $\Tp^{\mu\nu}=(1+a^2)^{1/2} \trho c^2  \uD^\mu\uD^\nu$. It has the ideal-fluid form with zero pressure and non-zero dimensionless enthalpy $w$ defined by $1+w=(1+a^2)^{1/2}$. Recall that the MHD electromagnetic field also has $T_{\rm f}^{\mu\nu}$ of an ideal-fluid form (\Eq~\ref{eq:Tf}), with effective pressure $\trho c^2\sigma/2$ and enthalpy $\trho c^2\sigma$.

As a result, one can replace $T^{\mu\nu}$ in \Eq~(\ref{eq:Q-}) with the effective $T_{\rm MHD}^{\mu\nu}=\trho c^2 h\uD^\mu\uD^\nu+P\eta^{\mu\nu}$ with the following definitions of $h$ and $P$: 
\beq
\label{eq:h}
  h=1+w+\sigma,
\eeq
\vspace*{-6mm}
\begin{subequations}
\beq
\label{eq:w_reg}
  w=\sqrt{1+a^2}-1, \qquad \frac{P}{\trho c^2}=\frac{\sigma}{2} \qquad ({\rm regular})
\eeq
\beq
\label{eq:w_stoch}
  w=\frac{3}{2}(\tg-1), \qquad \frac{P}{\trho c^2}=\frac{\sigma}{2}+\frac{w}{3}  \quad\; ({\rm stochastic})
\eeq
\end{subequations}
In the regular oscillation regime, the plasma makes no contribution to pressure $P$; the oscillations only increase the effective inertial mass of each particle by the factor $\tg=\sqrt{1+a^2}$. Unlike stochastic heating, coherent oscillations are not true heating. However, it is convenient to 
define effective enthalpy $w$ in both regimes, so that they have the common form of $T_{\rm MHD}^{\mu\nu}$.

The equations of mass and $Q^-$ conservation can now be stated as follows
\begin{align}
\label{eq:mass1}
   \partial_t\rho & = -\partial_\xi \left(\frac{\trho}{\kappa}\right), \\
\label{eq:Q1}
   \partial_t\left(\frac{\rho h}{\kappa}-\frac{P}{c^2}\right) &
           = -\partial_\xi\left(\frac{\trho h}{\kappa^2}\right) - \frac{n\dot{\Q}^-_e}{c^2}.
\end{align}
The term $\dot{\Q}^-_e=c(\dPe^t - \dPe^z)$ describes radiative losses per particle. It is determined by \Eqs~(\ref{eq:dE_dt_stream}), (\ref{eq:dP_dt_stream}) and \Eqs~(\ref{eq:dE_dt_stochastic}), (\ref{eq:dP_dt_stochastic}) in the regimes of regular oscillation and stochastic heating:
\begin{eqnarray}
\label{eq:dQ}
  \!\!\!\!\!\!  \dot\Q^-_e \! = \! \frac{c\sT U }{\gD \c^3} \!\! \times \!\!
   \left\{\begin{array}{lr}
       \vspace*{1mm}
              1+a^2                              &  \!\!\!\!\! {\rm regular} \\
      \displaystyle{\frac{5}{2}\tg^2}   &  \!\!\!\!  {\rm stochastic}
           \end{array} \! \right\} \! \approx \! \frac{c\sT U }{\gD\c^3} (1+w)^2. 
\end{eqnarray}
where $\sT=(8\pi/3)r_e^2$ is the Thomson cross section.

The radio wave does not contribute to \Eqs~(\ref{eq:mass1}) and (\ref{eq:Q1}), because it carries no mass and negligible $\Q^-$. The wave strength parameter $a(\xi)$ still enters the $\Q^-$ equation through $h=1+w+\sigma$ (since $a$ controls the fluid enthalpy $w$) and through $\dot{\Q}_e^-\propto \Uw\propto a^2$. As long as $a(\xi)$ is steady (not damped), it can be treated as a given function in the wave-fluid interaction problem, and \Eqs~(\ref{eq:mass1}) and (\ref{eq:Q1}) form a closed set for two unknowns $\rho(t,\xi)$ and $\c(t,\xi)$. The equations can be solved in the domain $0<\xi<T$ (i.e. inside the radio wave packet) with given initial conditions $\rho(0,\xi)$ and $\c(0,\xi)$, and boundary conditions $\rho(t,0)=\rhou$ and $\c(t,0)=\cu$.

\Eqs~(\ref{eq:mass1}) and (\ref{eq:Q1}) are Lorentz-invariant: they have the same form in any frame boosted along $z$.\footnote{Lorentz transformation of initial conditions is not, however, trivial, since hypersurfaces of simultaneity ($t=const$) are different in different frames.}
They can be solved e.g. in the frame where the upstream plasma is at rest: $\cu=1$ and $\rhou=\trhou$. The problem can also be solved in other frames where $\cu\neq 1$ and the upstream plasma has speed $\betau=(\cu^2-1)/(\cu^2+1)$, Lorentz factor $\gu=(\cu^2+1)/2\cu$, and density $\rhou=\gu\trhou$.


\section{Steady compression front}
\label{steady}

The pattern of fluid motion inside the propagating wave packet will be called below the ``front'' for brevity. Its structure is described by the profiles of $\rho(t,\xi)$ and $\c(\xi,t)$, which are governed by \Eqs~(\ref{eq:mass1}) and (\ref{eq:Q1}). In general, the front structure can evolve in time, as demonstrated in the next section. Here, we first show that there is a unique steady flow pattern, where all quantities depend on $\xi$ only, with no evolution with $t$. As verified in section~\ref{relaxation}, the steady state is an attractor for the front evolution. It is reached under two conditions: (1) the front propagation time is long enough for the flow pattern to relax to the steady state, and (2) the propagation time is short enough to neglect damping of the radio wave, so $a(\xi)$ can be treated as constant in time.

The steady front solution is found by setting $\left.\partial_t\right|_\xi=0$ and $\partial_\xi=d/d\xi$ in the MHD equations. Then, Equations~(\ref{eq:mass1}) and (\ref{eq:Q1}) give
\begin{align}
\label{eq:mass3}
  \frac{d}{d\xi} \left(\frac{\trho}{\kappa}\right) &=0  \quad \Rightarrow \quad \trho=\frac{\c}{\cu} \trhou, 
   \\
\label{eq:Q3}
  \frac{d}{d\xi}\left(\frac{h}{\kappa}\right) & = -\frac{\c \gD \dot{\Q}^-_e}{\me c^2}.
\end{align}
The corresponding lab-frame density $\rho=\gD\trho$ can be expressed using  
\beq
  \gD=\frac{\c^2+1}{2\c},
\eeq
so the solution $\trho/\trhou=\c/\cu$ gives
\beq
\label{eq:rho_steady}
  \frac{\rho}{\rhou}=\frac{1-\betau}{1-\beta} = \frac{\c^2+1}{\cu^2+1}.  
\eeq
The solution $\trho/\trhou=\c/\cu$ also implies $\sigma=\sigu\c/\cu$ and $d/d\xi(h/\kappa)=d/d\xi[(1+w)/\kappa]$. As a result, \Eq~(\ref{eq:Q3}) becomes
\beq
\label{eq:Q4}
    \frac{d}{d\xi}\left( \frac{1+w}{\kappa} \right) = - \frac{\sT U (1+w)^2}{\me c\, \c^2}.
\eeq
Note that the fluid magnetization $\sigma$ has dropped out. Remarkably, fluid motion and compression in a steady front are unaffected by the magnetic field, regardless of how strong it is. Integrating \Eq~(\ref{eq:Q4}) for $(1+w)/\c$ and using the cold upstream condition $w_{\rm u}=0$, we find 
\beq
\label{eq:steady_rad}
   \c=(1+q)(1+w)\cu, 
\eeq
where 
\beq
\label{eq:q}
   q(\xi) = \frac{\sT\Sigma(\xi)}{\me c^2\cu},  \qquad \Sigma(\xi)\equiv \int_0^{\xi} \Uw(\xi')c\,d\xi'.
\eeq
The effect of radiative losses on the fluid motion is controlled by the (Lorentz-invariant) dimensionless parameter $q$, which is determined by the column energy density of the radio wave packet, $\Sigma$. Losses weakly affect the front when $q\ll 1$.

In the regular oscillation regime with $q\ll 1$, the obtained steady-state solution for the compression front simplifies to 
\beq
\label{eq:steady_nr}
  \c=\cu\sqrt{1+a^2} \qquad \mathrm{(regular\; \& \; non\mathrm{-}radiative)}.
\eeq
In this limit, the solution agrees with the result of \cite{Sobacchi24}. This agreement is remarkable, since \cite{Sobacchi24} used a different approach to the steady-state compression front: they studied plasma in the frame moving with the group speed of a nearly monochromatic wave packet, and employed an expansion in Fourier harmonics, assuming that the upstream plasma ahead of the packet has $\tomB<\tom/a$.


\section{Relaxation to the steady state}
\label{relaxation}

A simple argument shows that the approach to the steady state takes the front a long propagation time if the plasma magnetization $\sigu$ is large. Consider the vacuum limit of $\rhou\rightarrow 0$, which corresponds to $\sigu\rightarrow\infty$.\footnote{Recall that we consider radio waves with frequencies exceeding the plasma Larmor frequency. In the opposite case, they would be MHD waves (where magnetized particles move with $\bD=E/B$), and $\rhou\rightarrow 0$ would correspond to force-free electrodynamics.}
Then, the background field must remain unchanged from its upstream state: in a vacuum, the electromagnetic wave goes through the background field without interaction, so no $\Ebg$ is generated. Thus, at $\sigu\rightarrow \infty$ the magnetized plasma must remain static and will not be compressed by the radio wave packet. The steady-state front with a large compression factor $C=\rho/\rhou$ still provides a consistent solution, but relaxing to it takes an infinite time, so it becomes irrelevant at $\sigu\rightarrow\infty$.

The same conclusion is reached by considering energy. Compression of the magnetically dominated plasma requires deposition of energy $\propto\sigu$, much greater than the energy of plasma particles that interact with the wave. The background field serves as a ballast coupled to the plasma. It increases the effective mass per particle from $m$ to $\sigu \me$, which makes it heavy and impossible to accelerate in the limit of $\sigu\rightarrow\infty$. An analogy can be drawn with absorption of a photon with energy $\E_{\rm ph}$ by a particle of mass $\sigma m$. Using the frame where the particle is initially at rest, one finds that the particle gains speed $\beta\approx \E_{\rm ph}/\sigma \me c^2\ll 1$ if $\sigma\gg\E_{\rm ph}/\me c^2$ (and $\E_{\rm ph}$ transforms mainly into the particle's internal energy).  

Similarly, one might expect that the interaction of a radio wave packet with the MHD fluid gives it a small bulk velocity, 
\beq
\label{eq:bD_in}
  \bD\sim \frac{w}{\sigu}\ll 1, 
\eeq
when the interaction generates fluid enthalpy $w\ll\sigu$. This would mean that energy deposition by the radio wave heats the fluid without significant bulk acceleration or compression.\footnote{A toy  problem of wave-plasma collision may be formulated treating the magnetized fluid as a single body carrying energy and momentum. A consistent formulation, however, requires some care: the MHD fluid would need to be confined by walls, which carry stress and energy to be included in the conservation laws.}
This estimate clearly differs from the steady-front solution (\Eq~\ref{eq:steady_rad}), which gives $\c\gg 1$ (i.e. $\bD\approx 1$) at $w\gg 1$ regardless of $\sigu$.

The estimate $\bD\sim w/\sigu$ may be reasonable at the beginning of the packet propagation through the plasma with $\sigu\gg w$; however, it fails to describe the gradual relaxation of the compression front to the steady state. When $\sigu\gg 1$, this relaxation takes a long time and involves a slow growth of the compression factor and the magnetic pressure in the front, as demonstrated below. Pressure gradients across the front play an essential role: they endow the propagating front with a ``memory'' of its earlier interactions with the fluid that has been disposed behind the front. The accumulating memory enables the front relaxation to the steady state. 

Below we demonstrate this relaxation by solving the full time-dependent MHD problem described by \Eqs~(\ref{eq:mass1}) and (\ref{eq:Q1}). We solve it first in the non-radiative regime (neglecting $\dot{\Q}_e^-$ in \Eq~(\ref{eq:Q1})) and then in the radiative regime. In the remainder of this section, we assume regular oscillations of particles in the radio wave. A similar relaxation trend holds with stochastic heating; however, this regime gives a huge enthalpy $w$ and relaxation would take a very long time, so steady fronts with stochastic heating do not form in FRBs (section~\ref{stochastic}).

\subsection{Evolution of a non-radiative compression front}
\label{non-radiative}

Before solving numerically the time-depending MHD equations, we make a heuristic estimate of the timescale $\trel$ for the front relaxation toward the steady state. 

First, note the time it takes a fluid element to cross the entire wave packet $0<\xi<T$:
\beq
\label{eq:tcross}
  \tcross=\int_0^T \frac{d\xi}{1-\bD} = \int_0^T \frac{\c^2+1}{2}\,d\xi. 
\eeq
In particular, in a steady state and in the regular oscillation regime, $\c^2=(1+a^2)\cu^2$.
Then, for a packet with $a_{\max}\gg 1$, one finds
\beq
\label{eq:tcross_steady}
   \tcross^\star\approx \int_0^T   [ a^2(\xi)\,\cu^2+1] \frac{d\xi}{2} 
   = \left(\frac{3}{8}a_{\max}^2\cu^2 + 1 \right)\frac{T}{2},
\eeq
where we assumed a wave packet $a(\xi)$ of the form given in \Eq~(\ref{eq:packet}). Hereafter, star indicates a steady state.

Now consider the momentum per particle carried by the MHD fluid inside the wave packet,
\beq
   \P_{\rm MHD}=\frac{T^{tz}_{\rm MHD}}{n c}=h\gD\bD\me c.
\eeq
It grows when the fluid receives momentum from the radio wave. Recall that the wave directly interacts only with the plasma particles, and the plasma shares the received momentum with the background electromagnetic field (through gyration about $\bB$). Therefore, the budget for the wave-fluid momentum exchange (per particle crossing the packet) is set by $T^{tz}_{\rm p}$ rather than $T^{tz}_{\rm MHD}$: 
\beq
  \Delta\P=\frac{T^{tz}_{\rm p}}{n c}=(1+w)\gD\bD\me c.
\eeq
When the fluid flow in the packet is far from the steady state, $\c\ll\cs$, the momentum flux in $\xi$ is non-uniform, $\partial_\xi F_{\rm mom}\neq 0$, and momentum comparable to $\Delta \P$ is retained inside the propagating front $0<\xi<T$, supporting $(\partial_t\c)_\xi>0$ and $(\partial_t\PMHD)_\xi>0$.

Thus, as long as $\c\ll\cs$, one may expect each particle crossing the wave packet to deposit $\sim\Delta\P$ on the timescale $\tcross$ toward future fluid momentum in the propagating front. Then, the timescale to grow $\PMHD$ is 
\beq
   t_{\P} \sim \frac{\P_{\rm MHD}}{\Delta \P}\tcross\sim \frac{h}{1+w}\tcross,
\eeq
where $h=1+w+\sigma$. Let us specialize to the frame where the upstream fluid is at rest ($\cu=1$). The steady front has $\c=\cs=1+w$ and $\sigma=\cs\sigu=(1+w)\sigu$. In this frame, 
\beq
   \tcross^\star \approx \frac{3}{16} a_{\max}^2T  \qquad (a_{\max}\gg 1),
\eeq
and the expected timescale to approach the steady state is 
\beq
\label{eq:trel}
  \trel\sim\frac{1+w+\cs\sigu}{1+w}\,\tcross^\star=(1+\sigu)\tcross^\star. 
\eeq

\begin{figure}[t]
\includegraphics[width=0.47\textwidth]{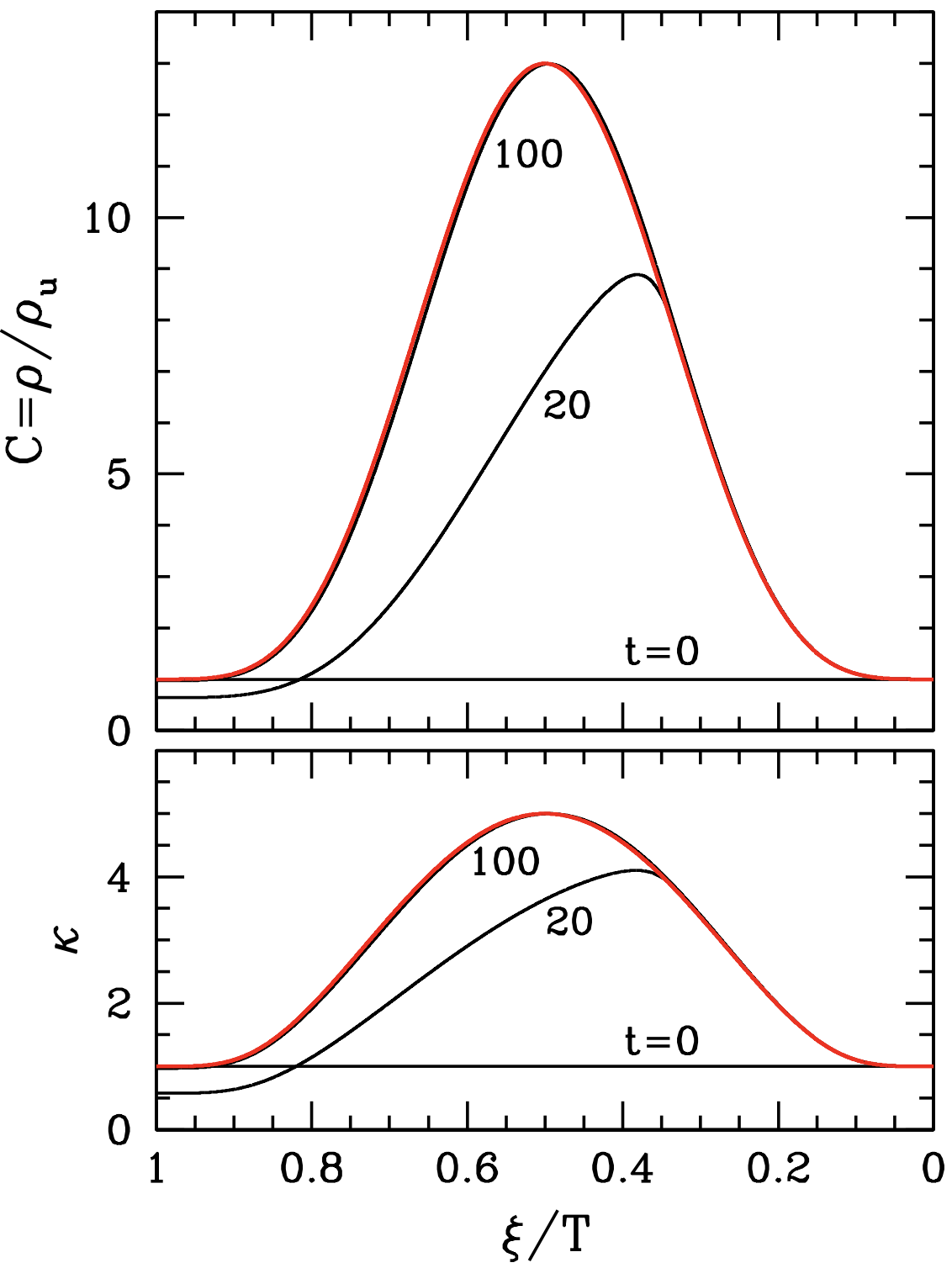}
\caption{Top: relaxation of the non-radiative compression front $C(\xi)$ toward the steady-state $C_\star(\xi)$ (red curve, \Eq~(\ref{eq:st})) from an initial state with a uniform static plasma, $C=1$ and $\c=1$. Black curves show snapshots of the front at times $t$ indicated next to each curve in units of packet duration $T$. The wave packet (\Eq~\ref{eq:packet}) has $a_{\max}=\sqrt{24}$, which gives  $w_{\max}=4$. The upstream magnetization is $\sigu=10$. Bottom: evolution of $\c$ (black) from the initial $\c=1$ toward $\cs$ (red) in the same simulation.}
\label{fig:relax}
\end{figure}

This simple estimate agrees with our numerical solution for the front evolution. We find the evolution by solving \Eqs~(\ref{eq:mass1}) and (\ref{eq:Q1}) for $\c(t,\xi)$ and $\rho(t,\xi)$ in a given wave packet $a(\xi)$ propagating in a background plasma with a given upstream magnetization $\sigu$. The fluid enthalpy inside the packet is given by \Eq~(\ref{eq:w_reg}). In sample models, we use the packet described by \Eq~(\ref{eq:packet}) with $a_{\max}=\sqrt{24}$, which corresponds to $w_{\max}=4$. As an initial state at $t=0$, we take a static plasma, $\c(\xi)=1$ and $\rho(\xi)=\rhou$, and follow the evolution of the flow pattern in the propagating wave packet until it relaxes to a steady-state. In the frame where the upstream plasma is at rest ($\cu=1$), the expected steady-state solution is
\beq
\label{eq:st}
   \cs=\sqrt{1+a^2},  \qquad  C_\star\equiv \frac{\rhost}{\rhou}=\frac{\cs^2+1}{2}=1+\frac{a^2}{2}.
\eeq 

The sample model with $\sigu=10$ is shown in Figure~\ref{fig:relax}. At the beginning of the simulation we observe that $\c(\xi)$ starts to grow from unity and develops a peaked profile --- the MHD fluid is accelerated in the leading part of the packet (at $\xi<\xi_\star$) and decelerated in the later part (at $\xi>\xi_\star$), before exiting behind the packet. The compression factor in the acceleration part is well described by the steady-state solution: 
\beq
  \c\approx \c_\star \;\; {\rm and} \;\; C\equiv \frac{\rho}{\rhou}\approx C_\star \quad {\rm at~} \xi<\xi_\star. 
\eeq
The maximum compression factor is reached at $\xi_\star$:
\beq
  C_{\max}=C(\xi_\star)\approx C_\star(\xi_\star). 
\eeq
The peak $C_{\max}$ gradually grows with time as its location $\xi_\star$ shifts toward the middle of the wave packet. 

\begin{figure}[t]
\includegraphics[width=0.47\textwidth]{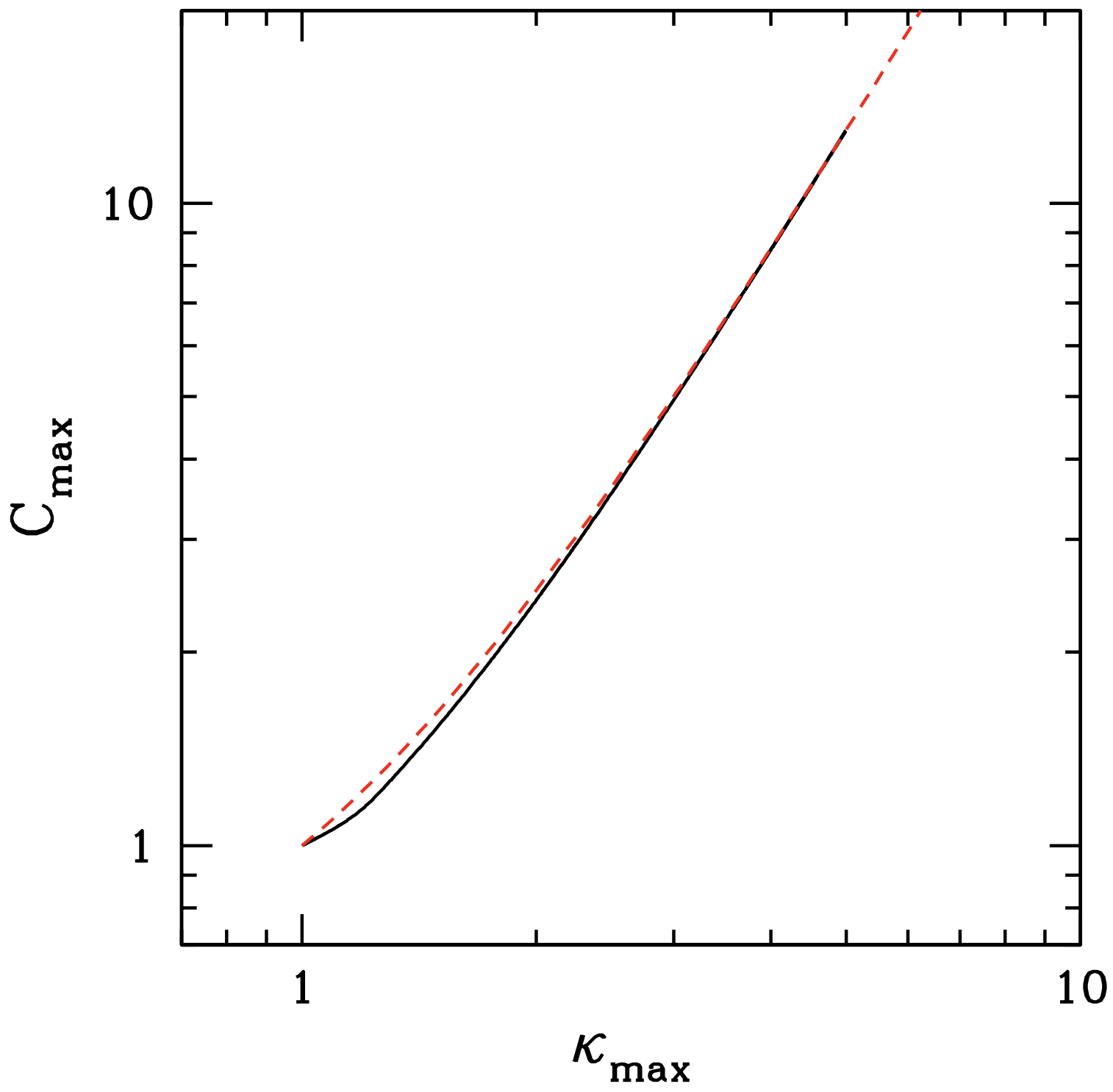}
\caption{Relation between $\c=\gD(1+\bD)$ and $C=\rho/\rhou$ (here measured at the peak inside the compression front) during relaxation toward the final steady state. Black curve shows the simulation result (the model with $\sigu=10$), and red dashed curve shows the relation $C=(\c^2+1)/2$ (equivalent to $\trho/\c=\rhou$ and uniform mass flux in $\xi$, $F_{\rm m}=\rho(c-v)$). Fluid density in the front closely tracks the steady-state solution $\trho=\c\rhou$ defined for an instantaneous profile of $\c(\xi)$, which slowly grows during relaxation. The correlated growth of $\c$ and $C$ stops when they reach $\c_{\max}^\star=5$ and $C_{\max}^\star=13$. Then, the fluxes of mass and $Q^-$ both become exactly uniform across the front, so a true steady state is achieved.}
\label{fig:c_C}
\end{figure}

When $\xi_\star$ reaches the middle of the wave packet, the entire compression front $\c(\xi)$ approaches $\cs(\xi)=\sqrt{1+a^2}$, which is an attractor for the flow pattern. In particular, the maximum of the $\c$-profile, $\c_{\max}$, eventually approaches the expected $\c_{\max}^\star=1+w_{\max}=5$. The maximum compression factor $C_{\max}=\rho_{\max}/\rhou$ approaches $C_{\max}^{\star}=1+a_{\max}^2/2=13$.

We also observe in the simulation that the gradual evolution of the front proceeds through a sequence of quasi-steady states: each snapshot of the front shows a nearly uniform mass flux in $\xi$: $F_{\rm m}=\rho(c-\vD)\approx const$, so $\rho(1-\bD)\approx \rhou$ and $C\approx (1-\bD)^{-1}$. Thus, the relation $C\approx (\c^2+1)/2$ is sustained during the evolution, i.e. $\c$ and $C$ grow in a correlated way (Figure~\ref{fig:c_C}).

\begin{figure}[t]
 \vspace*{0.5mm}
\includegraphics[width=0.47\textwidth]{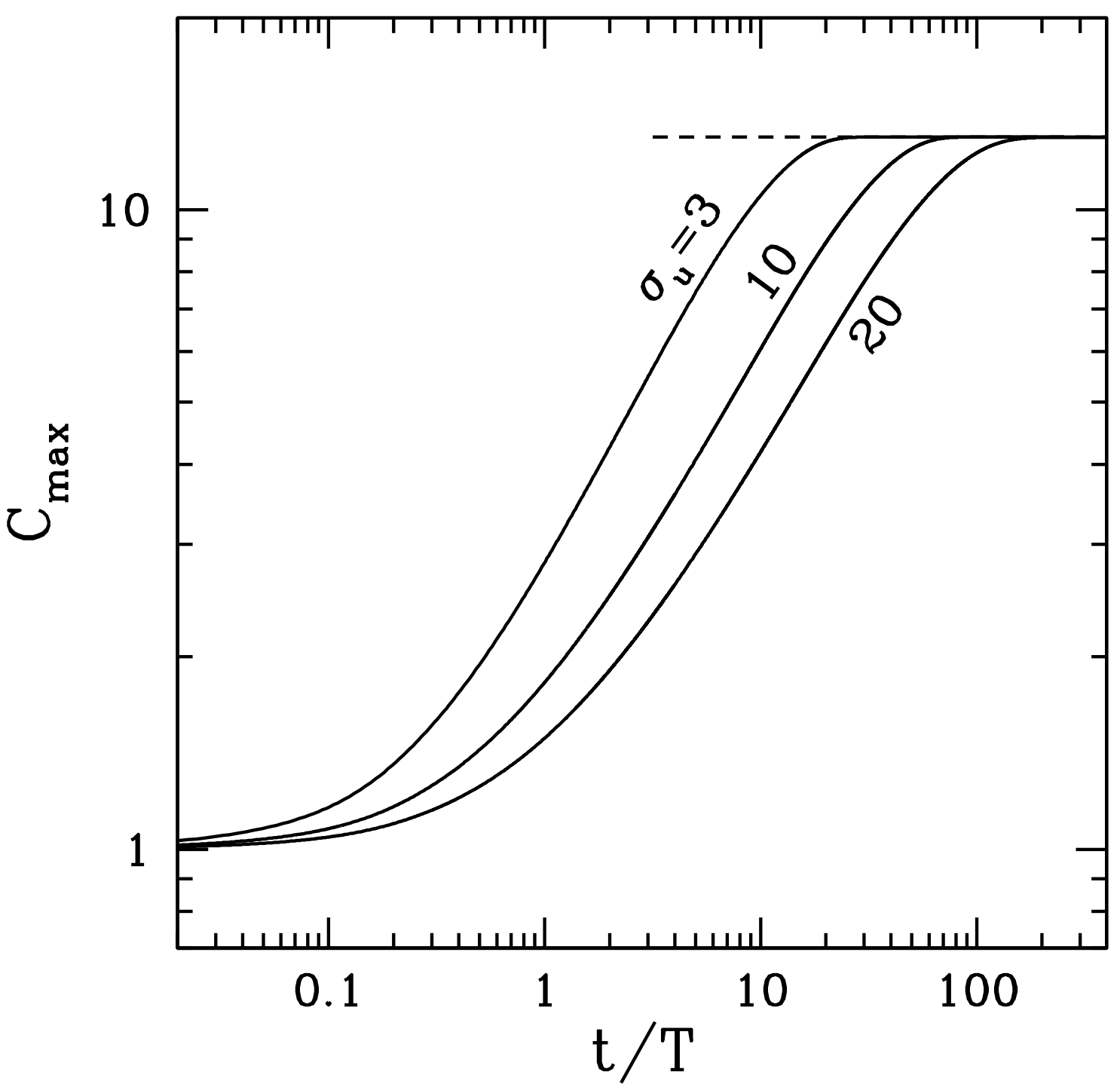}
\caption{Evolution of the compression peak $C_{\max}$ from the initial $C_{\max}=1$ to the steady-state $C_{\max}^\star=13$ (indicated by the horizontal dashed line) for the non-radiative fronts with $\sigu=3$, 10, and 20.}
\label{fig:Cmax}
\end{figure}

Evolution of the compression peak $C_{\max}(t)$ provides a simple way to quantify the front relaxation toward the final, exact steady-state. The evolution of $C_{\max}$ is shown in Figure~\ref{fig:Cmax} for $\sigu=3$, 10, 20. One can see that the timescale $\trel$  for approaching the steady-state $C_{\max}^\star=13$ increases with $\sigu$ in agreement with \Eq~(\ref{eq:trel}). 

We conclude that at any large but finite $\sigu$, the wave packet approaches the steady state on a long timescale $\trel\sim \sigu\tcross^\star$ (assuming negligible damping of the radio wave during this time). This conclusion may seem counter-intuitive: one might expect that the front has no memory on timescales longer than $\tcross$ since the plasma interacting with the wave packet is continually refreshed on the timescale $\tcross$. Nevertheless, the conservation laws require the propagating front to retain some memory of its earlier structure. Given enough time, the front gradually builds up the strong compression factor $\trho/\rhou=\c\approx 1+w$ across the entire front --- the pattern of the MHD flow in the wave is ``groomed'' toward the steady state. Remarkably, this gradual evolution leads to a strong magnetic field in the packet, as the plasma magnetization approaches $\sigma\approx (1+w)\sigu$. Thus, huge magnetic pressure develops in the propagating front while it finds a self-consistent steady pattern. At the same time, the final pattern of fluid motion in the wave packet is the same as found for radio waves propagating in an unmagnetized plasma. The magnetic pressure drops out from the steady-state solution $\c=\cs(\xi)$ because $\sigma/\c$ is exactly uniform across the compression front (section~\ref{steady}).

The relaxation of the compression front can be viewed as the evolution of a driven MHD wave in the plasma. The FRB drives a smooth compression wave of width $cT$, and it takes some time $t\sim t_\star$ for the wave pattern to find its equilibrium shape that propagates together with its driver without further change. Recall that in MHD compression tends to be redistributed across the plasma with Lorentz factor $\tg_{\rm s}\approx \sqrt{1+\sigma_{\rm eff}}$ (Appendix~\ref{vgr}). The compression pattern evolves due to the driving effect of the FRB and the communication of MHD information across the pattern, which occurs along characteristics ${\cal C}^-$ and ${\cal C}^+$. The ${\cal C}^\pm$ characteristics propagate in $z$ with speeds $\tilde{\beta}_\pm=\pm \tilde{\beta}_{\rm s}$ relative to the fluid, which itself moves with speed $\bD$ relative to a fixed lab frame. The ${\cal C}^\pm$ speeds in the lab frame $\beta_\pm$ are determined by the relativistic velocity transformation of $\tilde\beta_\pm$ from the fluid frame $\tKF$ to the lab frame. Instead of speeds $\beta_\pm$, it is convenient to use $\c_\pm\equiv\gamma_\pm(1+\beta_\pm)$, similar to $\c$ defined for the fluid speed $\bD$ in \Eq~(\ref{eq:kappa}). Then, the transformation between the frames takes the form
\beq
  \c_\pm=\tilde\c_\pm\c, \qquad \tilde\c_\pm\equiv\tg_\pm(1+\tilde\beta_\pm).
\eeq
Note that $\tilde\c_+\tilde\c_-=1$, since $\tilde\beta_+=\tilde\beta_{\rm s}=-\tilde\beta_-$. Magnetically dominated plasma with $\sigma_{\rm eff}\gg 1$ has $\tg_{\rm s}\approx \sqrt{\sigma_{\rm eff}}$, $\tilde\c_+\approx 2\tg_{\rm s}$ and $\tilde\c_-\approx (2\tg_{\rm s})^{-1}$.

The MHD characteristics ${\cal C}^\pm$ drift in $\xi$ with rate 
\beq
  \frac{d\xi_\pm}{dt}=1-\beta_\pm=\frac{2}{\c_\pm^2+1}. 
\eeq
Note that MHD information flows across the compression front in the direction of increasing $\xi$: $d\xi_\pm/dt>0$. As one can see in Figure~\ref{fig:relax}, the front evolution is basically the gradual spreading of the relaxed zone $\c=\c_\star$ from the leading edge $\xi=0$ toward larger $\xi$. The steady state is first established near the leading edge of the front where the relaxation time $\trel(\xi)$ is short because $\tcross(\xi)$ is short for a small leading part $\xi\ll T$. The middle part of the front takes longer to relax; however, its difference from the steady-state solution, $\c\neq \c_\star$, cannot perturb the already relaxed leading part, since MHD information does not spread to smaller $\xi$.

In a plasma with $\sigma_{\rm eff}\gg 1$, MHD information carried by ${\cal C}^+$ crosses the front much slower than ${\cal C}^-$:
\beq
  \frac{d\xi_+}{dt}\approx \frac{2}{\c_+^2} \approx \frac{1}{2\sigma_{\rm eff}\c^2}\ll  \frac{d\xi_-}{dt}.
\eeq 
Thus, the timescale for complete MHD relaxation is related to ${\cal C}^+$ propagation across the front rather than ${\cal C}^-$. The drift $d\xi_+/dt\approx 2/\c_+^2$ is slower than the plasma drift $d\xi/dt\approx 2/\c^2$ by the factor of $\tilde\c_+^2\approx 4\sigma_{\rm eff}$ (when using a lab frame where the plasma has $\c\gg 1$). Therefore, the timescale $t_+$ for MHD information to cross the compression front is $\approx 4\sigma_{\rm eff}$ times longer than the timescale for the plasma to cross the front:
\beq
  t_+\approx 4\sigma_{\rm eff}\tcross. 
\eeq
When the entire front approaches the steady state, $\c\approx\c_\star= \tg\cu$ implies $\sigma_{\rm eff}\approx \sigu$ (uniform across the front) and $t_+^\star\approx 4\sigu\tcross^\star$, where star indicates a steady state. This expression is similar to the heuristic estimate for relaxation time $t_\star$ given in \Eq~(\ref{eq:trel}), within the numerical factor of $4$.

\subsection{Evolution of a radiative compression front}
\label{radiative}

If $q\gg 1$, the fluid element moving through the wave packet experiences strong radiative losses. In this case, the radio wave continually deposits energy to offset the losses and keep  the average $\tg=\sqrt{1+a^2}$ \citepalias{Beloborodov22}. The deposited power per particle in the fluid frame approximately equals the average emitted power $\delta\tilde{\E}_e/\delta \tilde{t}$ (given by \Eq~(\ref{eq:dE_dt_stream}) with $\bD=0$), so the wave-fluid interaction at $q\gg 1$ is essentially a scattering process.

The scattering exerts a force on the fluid. The wave passes momentum with rate $\delta\tilde{\E}_e/c\delta \tilde{t}$, and the fluid radiates momentum with rate $\delta\tilde{\P}_e^z/\delta\tilde t$, so the net rate of momentum deposition into the MHD fluid is
\beq
\label{eq:dPsc_}
   \frac{\delta\tilde{\P}}{\delta \tilde{t}}
   =\frac{\,\delta \tilde{\E}_e}{c\delta \tilde{t}}-\frac{\delta\tilde{\P}_e^z}{\delta\tilde t}.
\eeq 
Transformation from the local fluid frame to a lab frame gives $\delta\P/\delta t=\delta\tilde{\P}/\delta \tilde{t}$ since $\delta t=\gamma\delta\tilde t$ and $\delta \P=\gamma\delta\tilde \P$ (from Lorentz transformation, using $\delta\tilde\E/\delta\tilde\P=\tilde{v}_{\rm D}=0$). Thus, \Eq~(\ref{eq:dPsc_}) gives the Lorentz-invariant force $\dot\P_{\rm sc}=\delta\P/\delta t$ applied to the MHD fluid by the scattering of the radio wave. Substitution of \Eqs~(\ref{eq:dE_dt_stream}) and (\ref{eq:dP_dt_stream}) (with $\bD=0$ in frame $\tKF$) yields
\beq
\label{eq:dPsc}
   \dot\P_{\rm sc}=\frac{\sT \Uw }{\c^2}(1+a^2).
\eeq

A fluid element crossing the front gains from scattering the following net momentum per particle:
\begin{align}
\nonumber
   \Delta\Psc &= \int_0^{\tcross} \dot\P_{\rm sc} \, dt=\int_0^T \frac{\dot\P_{\rm sc}\,d\xi}{1-\bD} \\
   &\sim \left(1+\frac{1}{\c^2}\right)\sT \Uw a^2 T,
\end{align}
with $a\sim a_{\max}$. This sets the budget for the fluid momentum gain per crossing time $\tcross$. The gained momentum is shared with the magnetic field frozen in the fluid, increasing $\PMHD$.

Similarly to the non-radiative case, one can make a heuristic timescale estimate for the growth of $\PMHD$ in the frame where the upstream plasma is at rest ($\cu=1$): 
\beq
\label{eq:tP_rad}
   t_\P\sim \frac{\PMHD}{\Delta\Psc}\,\tcross \sim \frac{h\c}{ q w^2} \,\tcross,
\eeq 
where $q$ may be evaluated using the column energy $\Sigma$ in the middle of the wave packet (\Eq~\ref{eq:q}). \Eq~(\ref{eq:tP_rad}) gives the timescale for approaching the steady state, $\trel=t_\P(\cs)$ if we substitute $\c\sim \cs =(1+q)(1+w)$ (\Eq~\ref{eq:steady_rad}) and $h_\star=1+w+\cs\sigu\approx\cs\sigu$. Interpolating between the limits of $q\ll 1$ and $q\gg 1$, we find
\beq
\label{eq:trel_rad}
    \trel\sim (1+q)\,\sigu \tcross^\star \qquad (\sigu\gg 1).
\eeq
This estimate may be compared with the results of numerical simulations presented below.

Front relaxation to the steady state in the radiative regime is found by solving \Eqs~(\ref{eq:mass1}) and (\ref{eq:Q1}) similarly to section~\ref{non-radiative}, but now keeping the term $n \dot\Q^-_e/c^2$. Using \Eq~(\ref{eq:dQ}) and $\Uw=\me\, \om^2 a^2/4\pi r_e$ (\Eq~\ref{eq:Uw}), we rewrite $\dot\Q^-_e$ as follows
\beq
\label{eq:R}
    \frac{\dot\Q^-_e}{\me c^2}  = {\cal R}\,\frac{a^2(1+a^2)}{\gD\c^3},
     \qquad   {\cal R}\equiv \frac{2r_e}{3c}\,\omega^2.
\eeq
The radiative parameter $q$ defined in \Eq~(\ref{eq:q}) is related to $\R$ by
\beq
  q(T)=\frac{\R}{\cu}\int_0^{T} a^2 d\xi = \frac{3}{8} a_{\max}^2 \frac{T\R }{\cu}.
\eeq
A characteristic value of $q$ in the middle of the front is $q(T/2)=q(T)/2$.

The value of parameter ${\cal R}$ depends on the choice of a reference frame, which can be parameterized by the upstream motion $\betau$ or $\cu$. Note that $\cu<1$ ($\betau<0$) in frames moving in the $+z$ direction relative to the upstream plasma. The wave frequency $\omega$ measured in a chosen frame scales as $\omega\propto \cu$, so ${\cal R}\propto \cu^2$. Note also that the packet width $T$ depends on the choice of a frame as $\cu^{-1}$ (since $\omega T=inv$), and the radiative parameter $q\propto T{\cal R}/\cu$ is invariant.

We use the following technical tricks to avoid two complications:

(1) For real FRBs, radiative losses may be important ($q>1$) only when $a \gg 1$, which also gives a large effective plasma enthalpy $w$ and a large compression factor $C$. We avoid simulating huge $w$ and $C$ by rescaling ${\cal R}$ to larger values, so that we can experiment with the radiative regime $q\gg 1$ at moderate $a$ accessible to simulations. In example models, we use the wave packet described by \Eq~(\ref{eq:packet}) with $a_{\max}=\sqrt{24}$ (which gives $w_{\max}=4$) and set the invariant radiative parameter $T{\cal R}/\cu=1/2$. This choice gives $q(T)=9/2$. A characteristic value of $q$ inside the pulse is $q\sim 3$.

(2) Both $\c$ and $\rho$ grow by a large factor during relaxation toward the steady state. The simulation more efficiently performed in a boosted frame where $\cu\ll 1$. This helps avoid ultra-relativistic motions of the accelerated fluid along the wave propagation direction (large $\c\gg 1$) that would slow down the simulation. We use the frame where $\cu=0.05$. When presenting the results, we transform the fluid parameters to the upstream rest frame (where $\cu=1$), so figures will show $\c$ and $C=\rho/\rhou$ measured in the upstream frame.

\begin{figure}[t]
\includegraphics[width=0.47\textwidth]{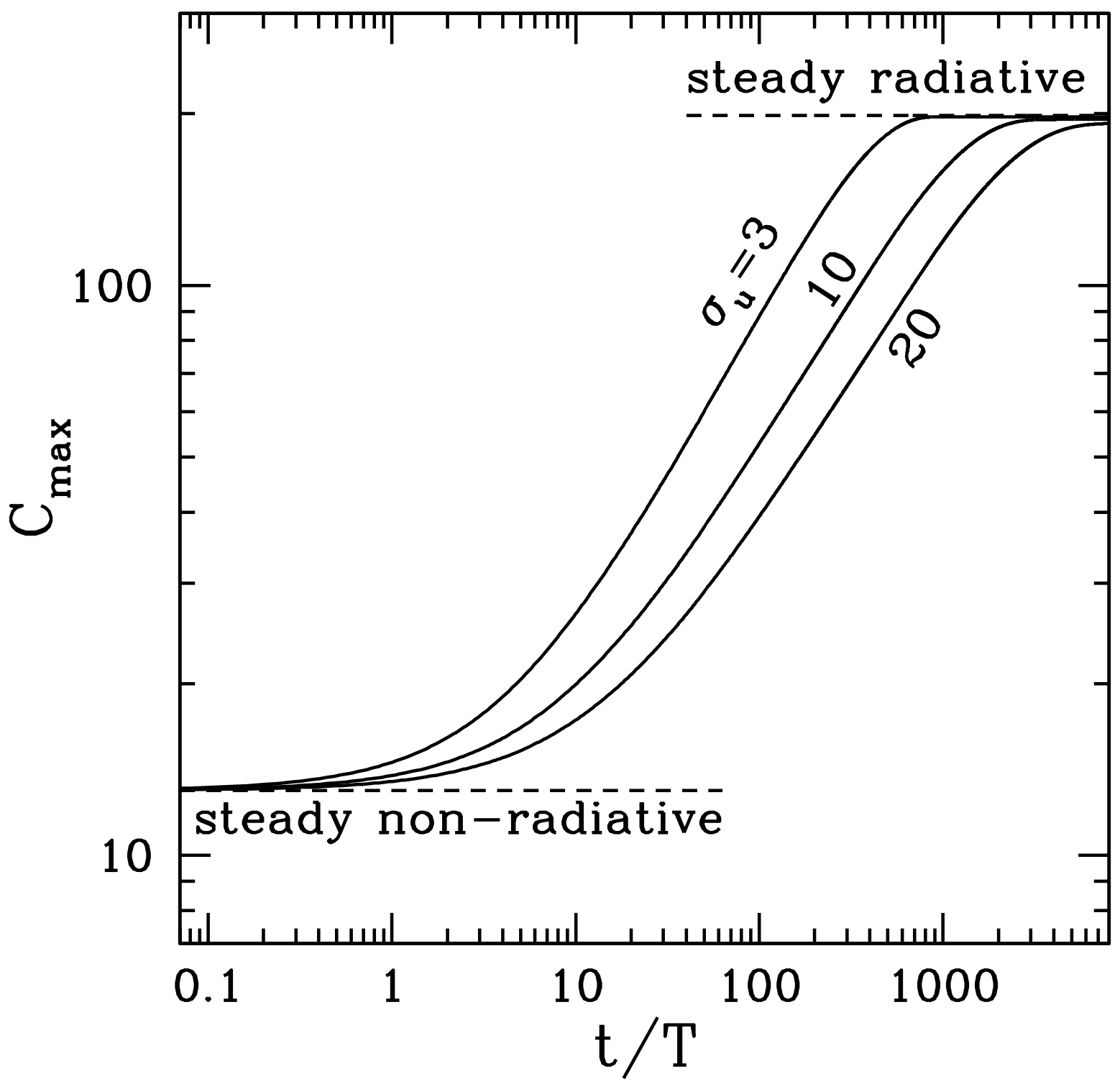}
\caption{Evolution of the compression peak $C_{\max}$ from the non-radiative steady state ($C_{\max}=C_{\max}^\star=13$ at $t=0$) after switching on radiative losses with $T{\cal R}/\cu=1/2$ (the radiative parameter ${\cal R}$ is defined in \Eq~(\ref{eq:R})). The front evolves toward the new steady state with $C_{\max}^\star\approx 200$. The evolution is shown for three simulations with upstream magnetizations $\sigu=3$, 10, 20. The initial and final states are independent of $\sigu$ and shown in Figure~\ref{fig:relax_rad}.}
\label{fig:Cmax_rad}
\end{figure}

\begin{figure}[t]
\includegraphics[width=0.47\textwidth]{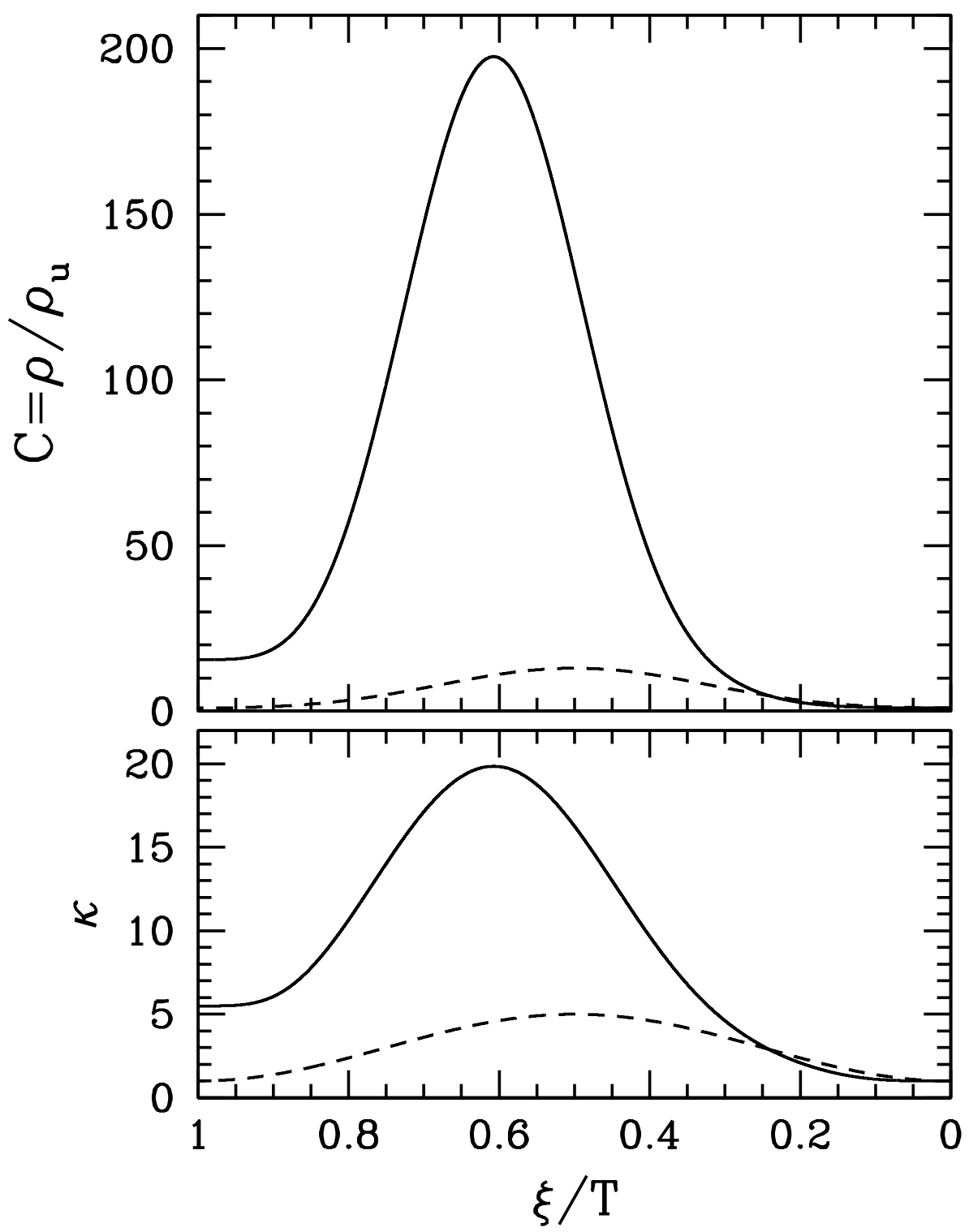}
\caption{Initial (dashed) and final (solid) profiles of the compression front in the three simulations shown in Figure~\ref{fig:Cmax_rad}. The initial state is the steady non-radiative front (Figure~\ref{fig:relax}, red curve). Radiative losses make the front evolve until it reaches the new steady state, which reproduces the analytical solution (\Eq~\ref{eq:steady_rad}).}
\label{fig:relax_rad}
\end{figure}

As an initial condition at $t=0$, we take the steady flow pattern in the non-radiative front. In the rest frame of the upstream plasma, it is given by $\c_0(\xi)=\sqrt{1+a^2(\xi)}$ and $C_0(\xi)=1+a^2(\xi)/2$, so the initial compression peak is $C_{\max}(0)=13$. In the presence of radiative losses, this flow pattern is no longer a steady state, and it immediately begins to evolve at $t>0$. Both $\c$ and $C$ grow and eventually relax to a new steady state. 

Figure~\ref{fig:Cmax_rad} shows the evolution of $C_{\max}$, and Figure~\ref{fig:relax_rad} shows the final profiles of $\c(\xi)$ and $C(\xi)$. One can see that $\c_{\max}$ grows from its initial value $1+w_{\max}=5$ by the expected factor $1+q\approx 4$, reaching the final $\c_{\max}\approx 20$, and $C_{\max}$ increases by the factor of $(1+q)^2$, from 13 to $\approx 200$. The observed relaxation is controlled by the chosen radiative coefficient $\R$, which determines $q(\xi)$; in particular, $q\approx 3$ at $\xi\approx 0.62\,T$, where the final $\c(\xi)$ is maximum. 

The simulations also confirm the expected scaling of the relaxation timescale $\trel\sim (1+q)\sigu \tcross^\star$ at $\sigu\gg 1$  (\Eq~\ref{eq:trel_rad}). The scaling $\trel\propto \sigu$ is seen in Figure~\ref{fig:Cmax_rad}, which shows three models with $\sigu=3$, 10, and 20. Note also that $\tcross^\star\propto \cs^2 \propto (1+q)^2$ and hence $\trel$ is expected to scale as $(1+q)^3$. We observe that the relaxation timescale is indeed increased by a large factor of $\approx (1+q)^3/2$ compared to the non-radiative front relaxation shown in Figure~\ref{fig:Cmax}.

As expected, the final steady state (Figure~\ref{fig:relax_rad}) is independent of $\sigu$: it is the same in all three simulations with $\sigu=3,10,20$. The observed final state satisfies \Eq~(\ref{eq:steady_rad}) with $\cu=1$ (since we view the results in the upstream rest frame). Note that in the non-radiative regime ($q\ll 1$) fluid comes back to rest as it exits behind the wave packet. By contrast, fluid behind the radiative front ($q\gg 1$) retains a large speed $\bD\sim 1$ ($\c\gg 1$). This is the result of irreversible momentum deposition by the scattering of the radio wave.


\section{Parameters of the wind}
\label{wind}

The obtained picture of the radio wave interaction with MHD fluid can now be applied to FRBs propagating through a radial wind carrying a transverse magnetic field, with some magnetization parameter $\sigu$. The wind serves as the upstream medium for the radio wave packet, and we begin with a brief summary of its parameters. They are denoted with subscript ``u'' for consistency of notation in this paper.

The wind Lorentz factor $\gu$ and magnetization $\sigu$ are estimated in \cite{Beloborodov20} (denoted there as $\Gw$ and $\sigw$). Their typical values are $\sigu\sim 10^2-10^3$ and $\gu\sim 10-30$ at radii $r\gtrsim 10^{11}\,$cm. The wind is also characterized by the particle outflow rate $\dN$ and power $\Lw$, which is dominated by the Poynting flux. Its isotropic equivalent (power per unit solid angle multiplied by $4\pi$) is
\beq
\label{eq:Lw}
   \Lw \approx c r^2 \Bu^2 = \eta \me c^2 \dN, \qquad \eta=\sigu\gu.
\eeq
A spinning magnetar emits a spindown wind with power
\beq
\label{eq:Lsd}
  \Lw\approx \frac{\mu^2\Omega^4}{c^3}\approx 5.8\times 10^{37}\,\mu_{33}^2 P^{-4}\;{\rm erg\,s}^{-1},
\eeq
where $\mu$ is the magnetic dipole of the magnetar and $P=2\pi/\Omega$ is its rotation period (in seconds). The spindown power gives a lower bound on $\Lw$, which can be enhanced during magnetar activity.

The wind forms at the light cylinder $\RLC=c/\Omega$ and its Lorentz factor grows as $\gu(r)\approx r/\RLC$ until it reaches $\eta^{1/3}$; at larger radii, $\gu$ grows logarithmically \citep{Beskin98,Lyubarsky01}. We will use a simple approximation:
\beq
\label{eq:wind}
  \gu\approx \left\{\begin{array}{cc}
            \vspace*{2mm}
          \displaystyle{\frac{r}{\RLC}} & \quad r<\eta^{1/3}\RLC \\
          \displaystyle{\left[\frac{\eta}{\ln 2} \ln\left(1+\frac{r}{\eta^{1/3}\RLC}\right)\right]^{1/3}} &\quad  r>\eta^{1/3}\RLC  \\
                            \end{array}\right.
\eeq
Our description of the wind is imprecise at small radii $r\sim \RLC$ for two reasons. (1) The wind magnetic field is assumed to be transverse to the radial direction. In fact, it has a radial component $B_r$ (which quickly decreases at $r>\RLC$). (2) The wind speed is taken equal to the drift speed. This approximation would be strongly violated for ordinary radio pulsars, where plasma outflows at $\RLC$ with a high Lorentz factor; however, it is reasonable for magnetars whose radiation field makes the outflow mildly relativistic \citep{Beloborodov20}. Our wind description should hold outside a few $\RLC$.

The parameter $\eta=\sigu\gu$ equals the plasma magnetization at the light cylinder,
\beq
\label{eq:eta}
   \eta=\sigma_{\rm LC}\sim \frac{\mu^2}{4\pi \RLC^3\N\me c^2}
   \approx  10^4\,\frac{\mu_{33}^2}{R_{\rm LC,10}^3\N_{37}},
\eeq
where $\N\equiv n r^3$ is the density parameter of the magnetosphere; its typical value is $\N\sim 10^{37}$ \citep{Beloborodov20}. The parameter $\eta$ and the Lorentz factor $\gu(r)$ determine the wind magnetization $\sigu(r)=\eta/\gu$.

Note that $\Lw$ determines the wind magnetic field in the static lab frame $\Bu(r)$  according to \Eq~(\ref{eq:Lw}). The corresponding magnetic field in the rest frame of the wind (ahead of the FRB wave packet) is $\tB_{\rm u}=\Bu/\gu$. It sets another important parameter --- the gyro-frequency of the wind plasma, 
\beq
  \tomB^{\rm u}=\frac{e\tB_{\rm u}}{\me c} 
  = \frac{e}{\gu\me c r}\left(\frac{\Lw}{c}\right)^{1/2},
\eeq
which should be compared with the FRB frequency. When viewed in the wind rest frame, the FRB frequency $\omega$ is redshifted to $\tom_{\rm u}=\omega/\cu$ where $\cu\equiv\gu(1+\betau)\approx 2\gu$. Then, we find
\beq
\label{eq:tomBu}
   \bup \equiv \frac{\tomB^{\rm u}}{\tomu} =\frac{\rb}{r},  \quad\;\, 
    \rb=\frac{2e\Lw^{1/2}}{\me c^{3/2}\omega}\approx 10^{11} \frac{L_{\rm w,37}^{1/2}}{\nu_9}\,{\rm cm}.
\eeq 

Below we examine the compression front driven by an FRB emitted at some radius $r$ and propagating outward through the wind. We first describe the front at large $r$, where plasma responds to the radio wave in the regular oscillation regime. Then, we consider FRBs emitted at smaller $r$, where stochastic heating occurs and leads to more dramatic effects, including damping of the FRB.


\section{Regular oscillation zone}
\label{regular}

Strong compression is expected at radii where the FRB wave packet has $a_{\max}>1$. The strength parameter $a_{\max}$ (\Eq~\ref{eq:a}) may be written as 
\beq
    a_{\max}=\frac{r_1}{r},
\eeq
where
\beq
\label{eq:r1}
   r_1 = \left(\frac{r_eL}{\me c\,\omega^2}\right)^{1/2} 
   \approx 1.6\times 10^{13}\,L_{42}^{1/2}\nu_9^{-1}\,{\rm cm},
\eeq
and $L$ is the peak luminosity of the FRB.

\subsection{Saturated compression ($r>r_\star$)}
\label{saturated}

One can quickly formulate a compression front model using the steady-state solution described by \Eqs~(\ref{eq:mass3}) and (\ref{eq:steady_nr}), which assumes regular oscillations of the plasma in the radio wave. The shape of the wave packet $a(\xi)$ then determines the profile of the proper compression factor in the front, $\tilde{C}\equiv \trho/\trhou$,
\beq
\label{eq:tc}
   \tC_\star= \frac{\c_\star}{\cu} = \sqrt{1+a^2},
\eeq
where star indicates that the solution is stated for the steady front. The corresponding compression of the wind density measured in the lab frame, $C\equiv\rho/\rhou$, is 
\beq
\label{eq:Csteady}
    C_\star(\xi)=\frac{\c_\star^2+1}{\cu^2+1} \approx 1+a^2,
\eeq
where $\c\equiv \gD(1+\bD)$ describes the fluid drift speed $\bD$, and we used $\cu\gg 1$ since the upstream medium is a relativistic wind. The maximum compression in the steady-state front is at the peak of the wave packet: $ C_\star^{\max}\approx 1+a_{\max}^2$.

The simple solution stated in \Eq~(\ref{eq:Csteady}) is valid only at sufficiently large radii $r>r_\star$ evaluated below. It does not describe compression fronts for FRBs at smaller radii for two reasons:
\\
(1) At $r<r_\star$, the compression front does not reach the steady state (section~\ref{unsaturated}). Then, the actual compression factor is well below its saturated value, $C\ll C_\star$. 
\\
(2) Compression promotes the transition to stochastic heating of plasma particles by the radio wave  (section~\ref{stochastic}). The plasma then receives much more energy (and hence much more momentum), which changes 
$\c$ and $C$.

\subsection{Unsaturated compression $(r<r_\star)$}
\label{unsaturated}

The saturated (steady-state) compression factor $C_{\max}^\star$ may be reached only when the wave packet has propagated for a time $t>\trel\approx (1+\sigu)\tcross^\star$ (section~\ref{non-radiative}). The magnetar wind has $\sigu\gg 1$ and moves with Lorentz factor $\gu\approx \cu/2\gg 1$. Then, the relaxation timescale $\trel$ may be expressed as
\beq  
  \trel \approx \frac{3}{4}\eta \gu a_{\max}^2 T,  
\eeq
where we used  \Eq~(\ref{eq:tcross_steady}) for $\tcross^\star$ and $\sigu\gu=\eta$.

Comparing $\trel$ with the FRB propagation time $t\approx r/c$ and using $\gu\approx \eta^{1/3}$ at $r\simgt \eta^{1/3}\RLC$, we find
\beq
   \frac{t}{\trel}=\frac{r^3}{r_\star^3}, 
\eeq
\beq
\label{eq:rsteady}
  r_\star\approx \left(\frac{ 2 r_e\E\eta\gu}{\me\omega^2}\right)^{1/3}
   \approx 1.5\times 10^{13}\,\frac{\E_{39}^{1/3}\eta_4^{4/9}}{\nu_9^{2/3}}\,{\rm cm}.
\eeq
Here, $\E$ is the FRB energy, which is related to the peak luminosity $L$ by
\beq
   \E=\frac{L}{a_{\max}^2}\int_0^T a^2 d\xi=\frac{3}{8}LT,
\eeq
where the numerical factor $3/8$ is found for the concrete shape of the radio wave packet given by \Eq~(\ref{eq:packet}). The expression for $r_\star$ assumes $a_{\max}^2(r_\star)>1$, i.e. it holds as long as $r_\star<r_1$. At the transition radius $r_\star$, we find
\beq
\label{eq:C*}
    C_{\max}^\star(r_\star)\approx  1+\frac{r_1^2}{r_\star^2}.
\eeq

At radii $r<r_\star$, the propagation time $t<\trel$ implies that the wave packet develops compression factors $C_{\max}(t)<C_{\max}^\star$ (Figure~\ref{fig:Cmax}). In this regime, $C_{\max}$ may be approximated by
\beq 
\label{eq:C_unsaturated}
   C_{\max}-1\approx (C_{\max}^\star-1) \left(\frac{t}{\trel}\right)^{\alpha}
   = a_{\max}^2 \left(\frac{r}{c\trel}\right)^{\alpha}.
\eeq
Numerical results in Figure~\ref{fig:Cmax} suggest $\alpha\approx 1/2$. The value of $\alpha$ depends on the assumed shape of the wave packet.


\section{Stochastic heating zone}
\label{stochastic}

\subsection{Boundary of the stochastic heating zone}

The transition from regular oscillations to stochastic heating is well defined on the $a$-$b$ plane (Figure~\ref{fig:trans}). The transition curve $b=\bs(a)$ is approximately described by \Eq~(\ref{eq:trans}). Even when $\bup\ll 1$ in the upstream plasma ahead of the FRB, the condition $b>\bs$ can be met inside the wave packet where the fluid is accelerated and compressed. Compression increases $\tomB$ and acceleration reduces $\tom$, so their ratio $b=\tomB/\tom$ is increased.

Fluid acceleration ($\c/\cu$) and compression ($\tC$ or $C$) are related by
\beq
\label{eq:c_C1}
   \tC\equiv\frac{\trho}{\trhou}\approx \frac{\c}{\cu}, \qquad C\equiv\frac{\rho}{\rhou}\approx\frac{\c^2+1}{\cu^2+1}.
\eeq
These relations are exact for a steady compression front (with $C=C_\star$ and $\c=\cs$) and also approximately hold for unsaturated compression, as demonstrated in Figure~\ref{fig:c_C} using the frame where the upstream plasma is at rest ($\cu=1$). \Eq~(\ref{eq:c_C1}) states the $\c$-$C$ relation viewed in other frames, where $\cu$ can have any value. 

The magnetic field is frozen in the fluid and compressed by the same factor as density: 
\beq
\label{eq:B}
  \frac{\tB}{\tB_{\rm u}} = \tC, \qquad \frac{B}{\Bu} = C.
\eeq
This implies  $\tomB\approx (\c/\cu)\,\tomB^{\rm u}$ while the FRB frequency $\omega$ transforms to the local fluid frame as $\tom=\omega/\c$, so one finds $b\approx (\c/\cu)^2\bup$. In the lab frame, the upstream medium (the wind) moves with $\cu\approx 2\gu\gg 1$. Then,  $(\c/\cu)^2\approx C$ and 
\beq
\label{eq:tomB2}
     b \approx  C \bup.
\eeq
Large compression $C$ inside the radio wave packet can lead to $b>\bs$, enabling stochastic heating even when $\bup\ll 1$. Therefore, the zone of stochastic heating $r<\rstoch$ extends far outside $\rb$ defined in \Eq~(\ref{eq:tomBu}).

\begin{figure}[t]
\includegraphics[width=0.47\textwidth]{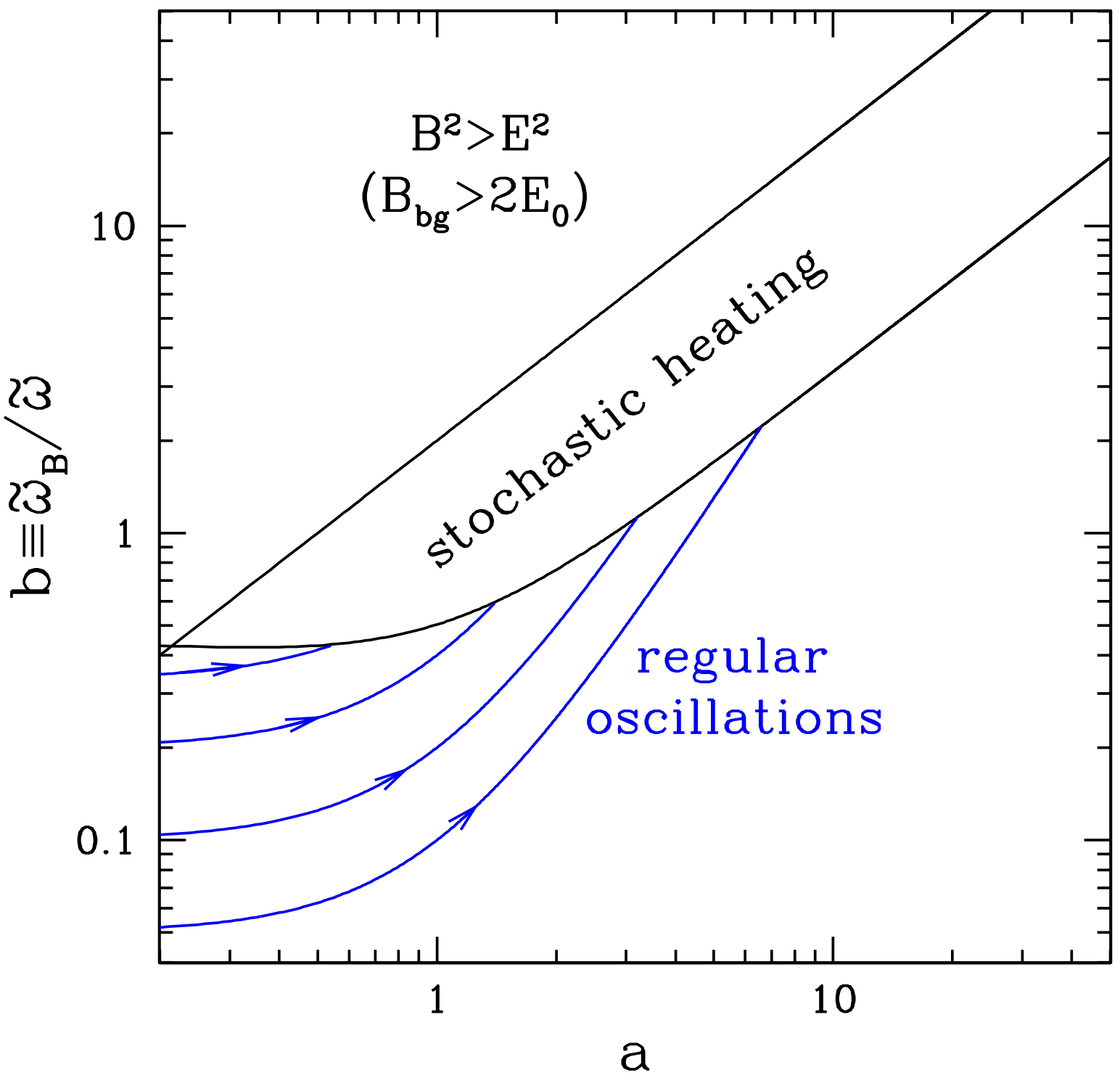}
\caption{Blue curves with arrows show the evolution of parameter $b\equiv\tomB/\tom$ in a fluid element exposed to an FRB. Fluid is initially in the upstream region where $a=0$ and $b=\bup=\rb/r$ (\Eq~\ref{eq:tomBu}). Parameter $b$ increases inside the radio wave packet in response to the rising $a$ (\Eq~\ref{eq:a_b}) and can eventually reach $\bs$, triggering stochastic heating. The four blue curves correspond to radii $r/\rb=3$, 5, 10, 20. Reaching the stochastic heating regime at large $r$ requires high FRB luminosities $L$ that give $\rstoch>r$ (\Eq~\ref{eq:rstoch}).}
\label{fig:trans_bu}
\end{figure}

The boundary $\rstoch$ is inside $r_\star$, and plasma compression by FRB at $r<r_\star$ occurs in the unsaturated regime: the peak $C_{\max}$ of $C(\xi)$ is reached at $\xi=\xi_\star<T/2$ (section~\ref{stochastic_heating}). The transition from regular oscillations to stochastic heating can only occur at $\xi<\xi_\star$ where $C(\xi)$ is rising. In this leading part of the wave packet, $C(\xi)$ tracks the steady-state $C_\star(\xi)$ (Figure~\ref{fig:relax}), so 
\beq
\label{eq:a_b}
 C\approx 1+a^2, \qquad b\approx (1+a^2)\bup.
\eeq
Stochastic heating is triggered where $b$ reaches $\bs(a)$:
\beq
\label{eq:trans_}
   (1+a^2)\bup\approx \bs(a).
\eeq
This equation defines $\astoch$ and $\Cstoch\approx 1+\astoch^2$ at the transition from regular oscillations to stochastic heating. 
Figure~\ref{fig:trans_bu} shows how fluid exposed to an FRB moves in the $a$-$b$ plane. At radii $r<\rstoch$, the trajectory $b(a)$ reaches $\bs(a)$, bringing the fluid to stochastic heating.

The transition point $b=\bs$ is reached if the wave packet has $C_{\max}\geq\Cstoch$. Using $C_{\max}(r)$ given in \Eq~(\ref{eq:C_unsaturated}), one can state the condition $C_{\max}\geq\Cstoch$ as 
\beq
\label{eq:stoch}
   a_{\max}^2\left(\frac{4r}{3cT\eta\gu a_{\max}^2}\right)^{\alpha} \geq \astoch^2,
\eeq
with $\alpha\approx 1/2$ suggested by our numerical models in section~\ref{non-radiative}. \Eq~(\ref{eq:stoch}) defines the zone $r\leq \rstoch$ where stochastic heating is triggered. Its boundary $\rstoch$ is found using the known $\bs(a)$ (Figure~\ref{fig:trans}), the wind parameters $\eta$, $\gu(r)$, and $\bup(r)$, and the known $a_{\max}(r)$ for a given FRB.

An approximate expression for $\rstoch$ may be obtained using \Eq~(\ref{eq:trans}) for $\bs(a)$. \Eq~(\ref{eq:trans_}) then gives
\beq
\label{eq:Cstoch}
   \astoch^2\approx \frac{1}{9\bup^2}-1=\left(\frac{r}{3\rb}\right)^2-1.
\eeq
The boundary $r=\rstoch$ of the stochastic heating zone (\Eq~\ref{eq:stoch}) satisfies
\beq
\label{eq:stoch1}
   \frac{r_1}{r}\left(\frac{4r}{3cT\eta^{4/3}}\right)^{1/2} \approx  \frac{r^2}{9\rb^2}-1,
\eeq
where we substituted $a_{\max}=r_1/r$ and $\gu\approx \eta^{1/3}$. For bright FRBs, $\rstoch\gg3\rb$ and \Eq~(\ref{eq:stoch1}) gives
\beq
\label{eq:rstoch}
  \rstoch\approx 10^{12} \frac{L_{\rm w,37}^{2/5} L_{42}^{1/5}}{\eta_4^{4/15}\nu_9^{6/5}T_{\rm ms}^{1/5}}\,{\rm cm}.
\eeq
This expression approximately holds for FRBs with $ L>10^{40}\,L_{\rm w,37}^{-1/2} \eta_4^{4/3} \nu_9 T_{\rm ms}$\,erg\,s$^{-1}$.

\subsection{Heating and radiative cooling in the wave packet} 
\label{heating_cooling}

At radii $r<\rstoch$, the FRB stochastically heats plasma particles to $\tg\gg a$, far exceeding the regular oscillation energy. In this regime, $\tg$ monotonically grows in the fluid element as it crosses the wave packet (Figure~\ref{fig:test_stoch}). The growth is well described by \Eq~(\ref{eq:tg_stochastic}) with the numerical coefficient $\chi\approx 0.8$:
\beq
\label{eq:stochastic1_}
      \tg^{4/3}\,\frac{d\tg}{d\xi} \approx 10\,
     \frac{r_e U \c^{2/3}}{\me \omega^{4/3}} \left(\frac{\tomBu}{\cu}\right)^{1/3},  
\eeq
where we substituted $a^2=4\pi r_e U/\me \omega^2$. Integration for $\tg(\xi)$ then gives
\beq
\label{eq:stochastic2}
  \tgstoch \approx \left[\frac{2 r_e \E}{\me c \omega r^2}\right]^{3/7} \! \! b^{1/7}
   \approx 10^5 \, \frac{\E_{39}^{3/7} L_{\rm w,37}^{1/14}\, C^{1/7} }{\nu_9^{4/7} r_{11}},
\eeq
where $\E=4\pi r ^2\Sigma=\int 4\pi r^2 U c\, d\xi$ is the energy of the radio wave (isotropic equivalent), and we substituted $b=C\bup$ (\Eq~\ref{eq:tomB2}) and used \Eq~(\ref{eq:tomBu}) for $\bup$. One can see that $\tg$ gained from stochastic heating exceeds the regular oscillation $\tg\sim a$ by a factor of $\sim 10^3$.

\Eq~(\ref{eq:stochastic2}) holds if radiative losses of the heated plasma are negligible. Energy losses in the fluid frame $\me c^2 \tdge=\Delta\tilde\E_e/\Delta \tilde t$ are given by \Eq~(\ref{eq:dE_stochastic}), which we rewrite as
\beq
   \tdge=\frac{3\sT \tUw}{2\me c}\, \tg^2.
\eeq
The losses offset the heating rate (\Eq~\ref{eq:stochastic}) when $\tg$ approaches 
\beq
  \tgrad\approx  \left(\frac{c}{r_e\tom}\right)^{\! 0.3} \! b^{0.1}
  \approx 10^4\, \frac{C^{1/4}\cu^{0.3}L_{\rm w,37}^{1/20}}{\nu_9^{0.4} r_{11}^{0.1}},
\eeq
where we used $b\approx C\bup$ and $\tom=\om/\c=(\om/\cu\sqrt{C})$.

In summary, stochastic heating gives
\beq
\label{eq:tgstoch}
   \tg = \min\{\tgstoch,\tgrad\}.
\eeq
The radiative regime $\tg=\tgrad$ occurs at radii that satisfy 
\beq
\label{eq:rad}
   r<\frac{6 \times 10^{11}}{\gu^{1/3}}\,\frac{\E_{39}^{10/21} L_{\rm w,37}^{1/42}}{C_2^{5/42}\, \nu_9^{4/21} } \,{\rm cm}.
\eeq

\subsection{Plasma trapping by the FRB}

When FRB propagation through the wind occurs in the zone of stochastic heating $r<\rstoch$, the plasma acceleration and compression turn out so large that it becomes trapped inside the radio wave packet. In general, large compression $C$ corresponds to large $\gD\approx C^{1/2}\gu$ (\Eq~\ref{eq:c_C1}) and implies a long timescale for the plasma to cross the wave packet $\tcross\sim \gD^2 T$ (\Eq~\ref{eq:tcross}). If $\tcross$ exceeds $t=r/c$, the plasma becomes trapped, surfing the radio wave. The ratio $\tcross/t$ may be written as 
\beq
   \frac{\tcross}{t}\sim \frac{cT\gD^2}{r} \approx \frac{\gD^2}{\gtrap^2} \approx \frac{C}{\Ctrap},
\eeq
where
\beq
  \gtrap\approx \left(\frac{r}{cT}\right)^{1/2}, \quad
   \Ctrap\approx \frac{r}{cT\gu^2} \approx \frac{30\,r_{11}}{T_{\rm ms}(\gu/10)^2}.
\eeq

In the regular oscillation zone $r>\rstoch$, trapping does not occur: $C_{\max}(r)$ is everywhere below $\Ctrap(r)$. Note that $C_{\max}$ is a decreasing function of radius while $\Ctrap$ increases with radius, so the ratio $C_{\max}/\Ctrap$ is largest at the inner boundary of the regular oscillation zone $\rstoch$. It is easy to verify that $C_{\max}(\rstoch)=\Cstoch<\Ctrap(\rstoch)$, where $\Cstoch$ is given by \Eq~(\ref{eq:Cstoch}).

By contrast, in the stochastic heating zone $r<\rstoch$, the FRB  deposits enormous energy and momentum into the plasma, resulting  in huge compression $C$. It would greatly exceed $\Ctrap$ if one assumes that the plasma crosses the wave packet.\footnote{A lower limit on the proper compression factor $\tilde{C}\approx C^{1/2}\approx\tilde\c/\tilde\c_{\rm u}$ may be estimated as $\tg/\sigu$, which gives $C\gg\Ctrap$.}
So, the plasma cannot cross the packet at $r<\rstoch$; it gets stuck inside the packet at some $\xi_{\rm trap}<T$.  

The trapping regime holds unless the FRB itself becomes strongly damped by its interaction with the wind. Damping allows the plasma to move through the wave packet while consuming its energy until $\xi_{\rm trap}$ approaches $T$ and the entire FRB loses its energy. The parameter space where this happens is evaluated below.


\section{FRB energy losses}
\label{damping}

Consider an FRB wave packet overtaking the wind and accelerating it from Lorentz factor $\gu$ to some $\gD$. Acceleration by the factor $\gu/\gD\approx\cu/\c\gg 1$ is accompanied by compression given in \Eq~(\ref{eq:c_C1}). 

Let the radio wave packet of energy $\E$ interact with $\Delta\N$ particles. Its energy loss $\Delta\E$ equals the sum of energy gained (and kept) by the MHD flow and energy emitted by the heated particles:
\beq
\label{eq:dEn}
   \Delta\E\approx \Delta\N \,\frac{T^{tt}_{\rm MHD}}{n}+\Delta\N\E_e,
\eeq
where $\E_e$ is the radiated energy per particle. In this relation, we have neglected the initial energy of the wind compared to its gain from the FRB. Using $T_{\rm MHD}^{tt}= \trho c^2 h\gD^2-P$ and dropping the term $P$ (it is smaller than the main term by the factor $\sim\gD^{-2}\ll 1$) one finds
\beq
   \frac{T^{tt}_{\rm MHD}}{n}\approx \me c^2\gD(1+w+\sigma).
\eeq

Radiative losses of a particle interacting with the radio wave packet are derived in Appendix~\ref{app:losses} in the regimes of regular oscillation and stochastic heating. For the MHD flow with $\gD\gg 1$ and $\tg\gg 1$ they simplify to
\beq
  \frac{\Delta \E_e}{\Delta t} \approx \frac{c\sT \Uw \tg^2}{2\c^2}.
\eeq
Then, we find
\beq
\label{eq:erad}
  \E_e = \! \int \! \frac{\Delta \E_e}{\Delta t} \, dt \approx \! \int \! \frac{c\sT \tg^2 U d\xi}{4}
  \approx \frac{\sT\tg^2\E}{16\pi r^2},
\eeq
where we used $dt=d\xi/(1-\bD)$ and $\c^2(1-\bD)=1+\bD\approx 2$. Note that \Eq~(\ref{eq:erad}) involves the FRB energy $\E$, assuming that the particle interacts with the entire radio packet, i.e. $\tcross\leq r/c$, which corresponds to $C\leq\Ctrap$.\footnote{A ``deep'' trapping regime $\tcross\gg r/c$ is possible if the plasma is trapped in a small leading part of the wave packet at $\xi_{\rm trap}\ll T$. In this case, $\E$ in \Eq~(\ref{eq:erad}) should be replaced by the part of the FRB energy $\E_{\rm trap}$ contained in the leading part $0<\xi<\xi_{\rm trap}$. Self-consistent $\xi_{\rm trap}$ and $\E_{\rm trap}$ depend on the packet shape. 
\label{foot:deepC}}

Using $\gD/\gu\approx  \sqrt{C}\gg 1$ and $\sigma/\sigu=\trho/\trhou\approx \sqrt{C}$ (\Eq~\ref{eq:c_C1}), we obtain
\beq
\label{eq:loss3}
    \frac{\Delta \E}{\Delta \N \me c^2 \gu} \approx \sqrt{C}\, w + C\sigu 
    + \frac{\sT\E\, \tg^2}{16\pi r^2\me c^2\gu}.
\eeq
As the FRB propagates to a radius $r$, it interacts with $\Delta\N$ wind particles given by
\beq
\label{eq:dN}
  \Delta \N \approx (1-\betau)\frac{r}{c}\dN\approx \frac{2r\Lw}{\me c^3\eta\cu^2}.
\eeq
\Eqs~(\ref{eq:loss3}) and (\ref{eq:dN}) determine the FRB energy loss $\Delta\E$, expressed as a function of the compression factor $C$. It is easy to verify that losses are negligible, $\Delta\E\ll \E$, in the zone of regular heating $r>\rstoch$. Thus, FRB damping in the wind may only occur in the zone of stochastic heating, $r<\rstoch$. Then, $w\approx (3/2)\tg$ and \Eq~(\ref{eq:loss3}) gives
\beq
\label{eq:loss4}
   \Delta \E \approx \frac{r\Lw}{2c\,\eta\gu}\left[\frac{3}{2}\tg \sqrt{C}+\sigu C + \frac{\sT\tg^2\E}{16\pi r^2 \me c^2\gu}\right].
\eeq
For an FRB with a given energy $\E$, we find $\tg(r,C)$ from \Eq~(\ref{eq:tgstoch}) at radii $r<\rstoch$, and then find $\Delta\E(r,C)$ from \Eq~(\ref{eq:loss4}).

\begin{figure*}[t]
\includegraphics[width=0.482\textwidth]{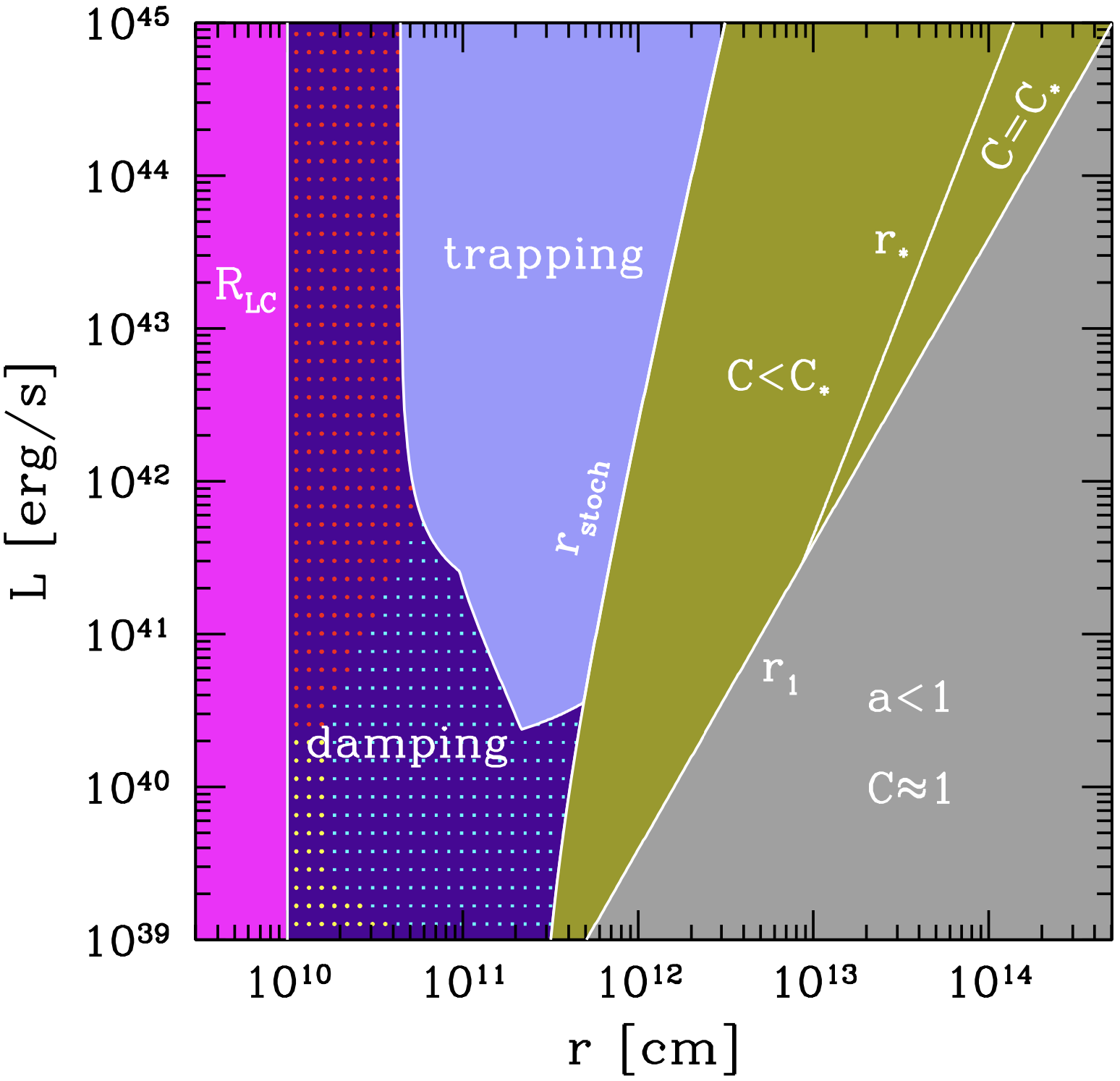}
\hspace*{4mm}
\includegraphics[width=0.482\textwidth]{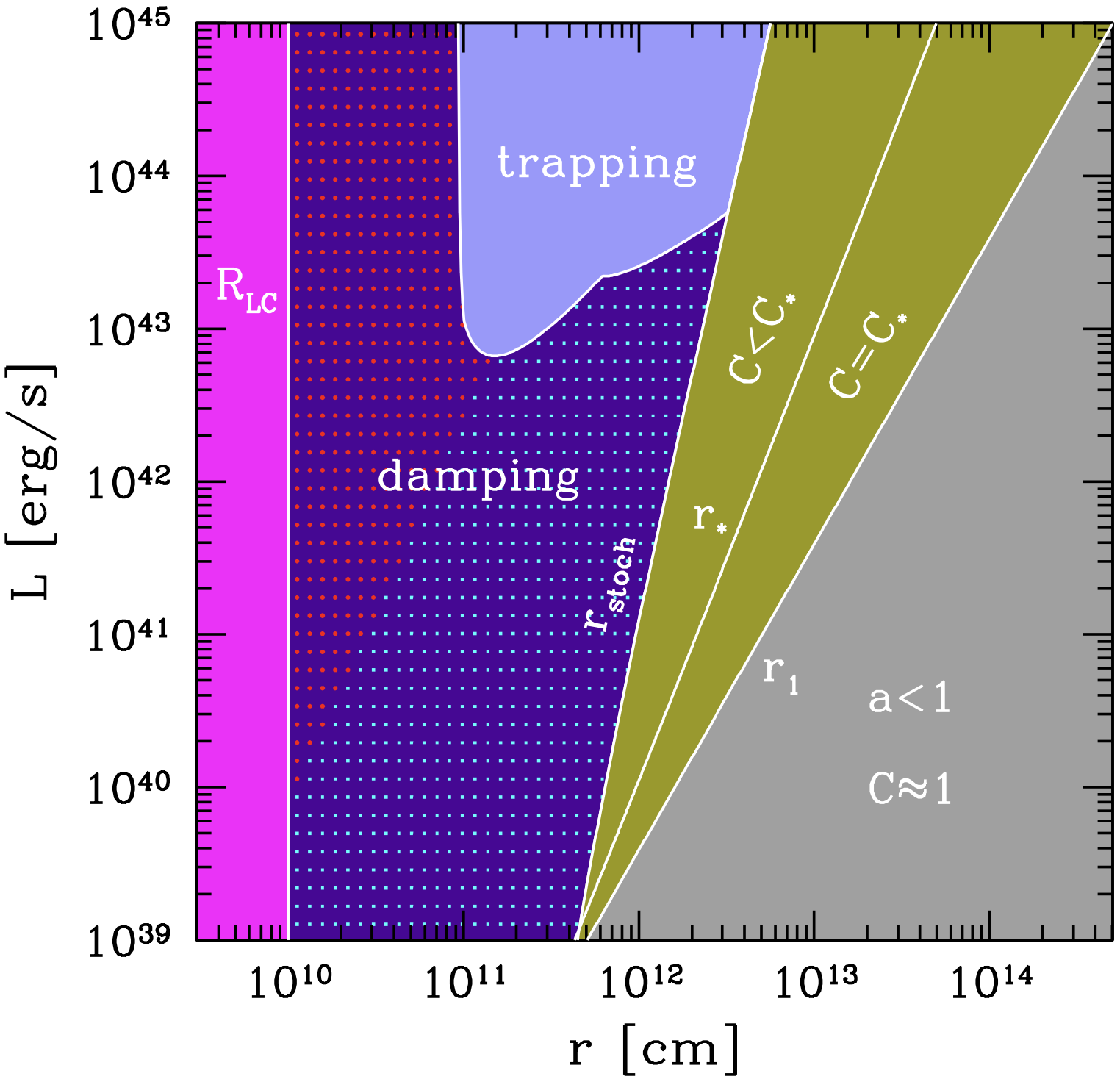}
\caption{Regimes of FRB propagation through the wind ($r>\RLC$), shown on the $r$-$L$ plane, for FRBs with frequency $\nu=1$\,GHz and duration $T=1$\,ms. The two plots assume the wind energy parameter $\eta=10^4$ (left) and $\eta=10^3$ (right). Magenta region $r<\RLC$ indicates the closed magnetosphere; FRBs emitted in this region are known to experience strong radiative damping \citep{Beloborodov24}. FRBs are also damped in the wind zone $r>\RLC$ studied in the present paper, in particular in the deep purple region. Their energy is consumed by the stochastic heating and compression of the wind plasma, with radiative losses.  Colored dots indicate the dominant damping channel: radiative losses $\Erad$ (red dots), energy kept by the plasma $\Eplasma$ (cyan dots), and energy spent to compress the wind magnetic field $\Emag$ (yellow dots). Stochastic heating occurs at $r<\rstoch$. In the light purple region, FRBs trap the wind plasma, carrying it in a leading part of the wave packet and avoiding  significant energy losses. In the olive zone, $\rstoch<r<r_1$, FRBs induce regular particle oscillations with $\tg\gg 1$ and drive a compression wave through the wind, with a decreasing compression factor $C$. At radii $r>r_\star$, the compression wave is locally quasisteady and $C=C_\star\approx 1+a^2$. The FRB strength parameter $a$ drops below unity outside $r_1$ (grey region).}
\label{fig:rL}
\end{figure*}

The FRB is strongly damped if $\Delta\E$ approaches $\E$. Damping begins in the leading part of the FRB $\xi\ll T$, as the plasma entering the radio wave packet consumes its local energy $\E(\xi)\sim 4\pi r^2 \Uw(\xi)\, c\,\xi$ and makes the deep trapping regime $C\gg\Ctrap$ unsustainable. Damping succeeds in killing the FRB when it enables the plasma to keep going in $\xi$ and crossing the entire wave packet, which corresponds to $\Delta\E\sim\E$ and marginal trapping, $C\approx \Ctrap$. Therefore, the condition for marginal damping is 
\beq
\label{eq:damping}
  \frac{\Delta\E(r,\Ctrap)}{\E}\approx 1.
\eeq 
It defines the boundary of the parameter space where FRBs are strongly damped. One can show that $\Delta\E(r,\Ctrap)/\E$ is a decreasing function of $\E$, so strong damping occurs for FRBs with energies $\E$ below a critical value $\E_{\rm esc}(r)$ defined by \Eq~(\ref{eq:damping}), and the corresponding critical luminosity is $\Lesc\sim \E_{\rm esc}/T$. Figure~\ref{fig:rL} shows the damping region on the $r$-$L$ plane, where $L$ is the peak luminosity of the FRB. In this example, we set the FRB duration $T=1$\,ms and assumed the shape of the radio wave packet stated in \Eq~(\ref{eq:packet}), which gives $\E=(3/8)LT$. 

It is useful to examine more closely the loss $\Delta\E$ stated in \Eq~(\ref{eq:loss4}). The three terms in the square brackets correspond to three contributions,
\beq 
   \Delta\E=\Eplasma+\E_{\rm mag}+\E_{\rm rad},
\eeq
where $\Eplasma\propto\tg$ is the energy kept by the plasma particles, $\Emag\propto\sigu$ is the energy used to compress the wind magnetic field, and $\Erad\propto\tg^2$ is the radiative losses. The largest of these three contributions defines the main damping channel for the FRB; it is indicated in Figure~\ref{fig:rL}. It is easy to see that $\E_{\rm rad}>\Eplasma$ if $\tgrad<\tgstoch$; in this regime, the plasma radiates more energy than it keeps. This occurs at radii that satisfy the approximate condition stated in \Eq~(\ref{eq:rad}) with $C=\Ctrap$.


\section{Summary}
\label{discussion}

The basic physics problem studied in this paper is how a strong ($a\gg 1$) electromagnetic wave packet accelerates an MHD fluid. This problem has important applications to FRBs that propagate from the central engine (a neutron star) through the surrounding medium. For weak electromagnetic waves ($a\ll 1$), the acceleration effect is well known as the radiation pressure force, used in astrophysics e.g. to define the Eddington limit. For strong waves, the acceleration effect is known in unmagnetized plasmas; the plasma then becomes compressed in the wave packet by the factor $C\approx 1+a^2$. Interaction of strong waves with magnetized plasmas was not sufficiently understood. This paper presented a new approach to this problem, a physical picture of the interaction, and sample numerical solutions. Two facts help solve the problem: (1) the wave strength parameter $a$ controls the plasma enthalpy as shown in section~\ref{heating}, and (2) quantity $\Q^-=\E-c\P$ (where $\E$ is energy and $\P$ is momentum) is conserved for the MHD fluid---the radio wave deposits energy and momentum, but does not deposit any significant $\Q^-$.

A first estimate would suggest that plasma with a sufficiently large magnetization $\sigma$ is not strongly compressed by the radio wave, since the deposited momentum is proportional to the small plasma density while compression requires work to be done against the large background magnetic pressure (\citealt{Beloborodov21b}; see also section~\ref{relaxation}). The actual solution demonstrates that this expectation holds initially, when the wave packet begins to interact with the plasma. If the propagating radio wave is given sufficient time, it gradually develops a much larger compression factor $C$. After time $\trel\propto C\sigma$ the compression front relaxes to a steady state with $C=C_\star\approx 1+a^2$, which is independent of $\sigma$. 

Besides the peculiar gradual evolution of the propagating compression front, the background magnetic field brings another special feature: it enables powerful stochastic heating of the plasma by the radio wave, in particular when compression increases the local gyro-frequency. The heating boosts momentum deposition, further enhancing acceleration and compression of the MHD fluid in the wave packet.

We have used the obtained results to examine FRB interaction with a magnetized wind flowing from the neutron star. Our conclusions are summarized below and illustrated in Figure~\ref{fig:rL}.

FRBs have an enormous effect on the wind medium when they propagate at radii $r<\rstoch$. They heat the wind up to an average Lorentz factor $\tg\sim 10^3 a$ (measured in the fluid rest frame), and accelerate it to a bulk Lorentz factor exceeding $\gtrap\sim (r/cT)^{1/2}$, which implies trapping of the wind plasma inside the radio wave packet. The energy per particle in the heated plasma $\gtrap\tg\me c^2$ (measured in the static lab frame) is in the TeV range. Thus, the FRB-wind interaction converts the radio-wave energy to high-energy particles, which experience radiative losses. A fraction of FRB energy is also spent to compress the wind magnetic field. FRBs lose a small fraction of their energy if they are emitted into the wind at $r\simgt 10^{11}$\,cm (Figure~\ref{fig:rL}).

The FRB-wind interaction changes at larger radii: at  $r=\rstoch\sim 10^{12}$\,cm (\Eq~\ref{eq:rstoch}) there is a sharp transition from stochastic heating to regular particle oscillations in the radio wave, with a moderate Lorentz factor $\tg\sim a$. FRB energy losses are negligible at $r>\rstoch$. FRBs emitted in this zone still induce strong bulk acceleration of the wind plasma, but no longer trap it, and continue to drive an MHD compression wave in the wind. The compression wave becomes locally quasisteady at $r>r_\star\sim 10^{13}L_{42}^{1/3}$\,cm (\Eq~\ref{eq:rsteady}), with compression factor following the simple expression $C_\star\approx 1+a^2$. In this region, our results agree with \cite{Sobacchi24}. 

It should be emphasized that we find the compression factor $C\ll C_\star$ at $r\ll r_\star$, where the timescale for the compression front relaxation toward $C_\star$ exceeds the FRB propagation time $r/c$. Therefore, our results do not support the suggestion of \cite{Sobacchi24} that FRBs emitted at $r\lesssim 10^{12}$\,cm lose energy to the huge compression $C_\star$ of the magnetized wind.\footnote{Note also that energy consumption implied by $C_\star\approx 1+a^2$ is reversible: when the MHD fluid exits behind the wave packet (where $C_\star$ comes back to unity) it returns energy to the FRB.} 

Our results still imply a significant constraint on the FRB emission radius (Figure~\ref{fig:rL}). It is caused by the enormous stochastic heating of the wind exposed to the FRB at small $r$, which leads to catastrophic energy losses of the FRB. As a result, the previously found damping zone $r<\RLC$ \citep{Beloborodov21b,Beloborodov24,Golbraikh23} is now extended to $r_{\rm damp}\sim 10^{11}$\,cm. This will challenge any FRB scenarios that place radio emission at $r<10^{11}$\,cm and rely on its escape through the magnetar wind. The exact $r_{\rm damp}$ depends on the wind parameter $\eta=\Lw/\dN \me c^2$, which may vary around $\eta\sim 10^4$. We find that $r_{\rm damp}$ slowly decreases with increasing $\eta>10^4$. FRB damping is particularly strong in winds with $\eta<10^4$.

The constraint on the emission radius supports the explosion scenario for FRBs. In all cases, the central engine must be the inner magnetosphere ($r\simlt 10^8$\,cm), which possesses sufficient energy to power the observed burst. The energy liberated in the inner magnetosphere must be relocated  outside $r_{\rm damp}$ before the FRB is released into the surrounding medium. This is accomplished by sudden ejection of a macroscopic (ms-long) electromagnetic pulse, which expands to large radii. The ejected pulse is capable of producing FRB emission by two known mechanisms. (1) The pulse propagation through a preexisting current sheet near $\RLC$ results in small-scale perturbations imprinted in the ejecta. These perturbations may be released as GHz radio waves when the ejecta expands to a large radius \citep{Lyubarsky20}. (2) FRB is emitted by synchrotron maser in a relativistic shock created by the expanding ms electromagnetic pulse \citep{Beloborodov17b,Beloborodov20}. Such FRBs avoid damping because GHz emission occurs at $r\sim 10^{13}-10^{14}$\,cm, well outside $r_{\rm damp}$. The relativistic shock was also proposed to produce radio waves via mode conversion when it propagates in a medium with small-scale fluctuations \citep{Thompson23}.

In addition to plasma acceleration and compression by FRBs, the ambient medium may be strongly affected by self-modulation instabilities of the radio wave \citep{Sobacchi23}, investigated so far for waves with $a<1$. Instabilities in the wind zone with $a>1$ are poorly understood and will require future work. 

After this work was finished, a paper by \cite{Sobacchi25} appeared on arxiv, discussing conditions for stochastic heating and its interpretation as synchrotron absorption calculated by \cite{Lyubarsky18}. Within a  numerical factor, their conditions are consistent with both Figure~3 of the present paper and the condition $\gamma\gtrsim\gamma_\star$ stated in \citetalias{Beloborodov22}. Links between stochastic heating (\Eq~\ref{eq:stochastic}) and the absorption rate calculated by  \cite{Lyubarsky18} are explained in Appendix~\ref{stochastic_orbit}.

\medskip

I thank the Kavli Institute for Theoretical Physics for hospitality during the plasma astrophysics program in summer 2025, where this work (including Figures~3 and 9) was presented  and discussed with the program participants. I thank Yuri Lyubarsky for drawing my attention to his paper on synchrotron absorption of strong waves, which stimulated the discussion in Appendix~\ref{stochastic_orbit}. The author acknowledges support by NASA grants 21-ATP21-0056 and 80NSSC24K0282, NSF grant AST-2408199, and Simons Foundation grant 446228. This work was also facilitated by Multimessenger Plasma Physics Center (MPPC) grant PHY-2206609.


\appendix

\section{Speed of the radio wave packet}
\label{vgr}

For FRBs propagating through the wind outside the magnetosphere (section~\ref{wind}), the relevant propagation regime is such that particle oscillation in the radio wave occurs faster than Larmor rotation in the local magnetic field: $\tom>\tomL$ (where tilde indicates that the quantity is measured in the local fluid frame $\tKF$). Then, the magnetic field weakly affects the FRB propagation speed. The plasma response to the radio wave (i.e. the excited electric current)  depends on the local $e^\pm$ density $\tn$ and the effective particle mass $\tg \me$, where $\tg$ is the average particle Lorentz factor in the fluid frame. The radio wave then satisfies the standard dispersion relation, which can be stated in the fluid frame as
\beq
    \tom^2(\tilde k)\approx c^2\tilde{k}^2+\tom_{\rm p}^2 \qquad (\tom\gg\tomL),
\eeq
where
\beq
     \tom_{\rm p}^2=\frac{4\pi e^2 \tn}{\tg\me}. 
\eeq
The FRB wave packet propagates relative to the plasma with group speed $\tilde v_{\rm gr}=d\tom/d\tilde k=c^2\tilde k/\tilde\omega$. It can also be characterized by Lorentz factor
\beq
   \tg_{\rm gr}=\left(1-\frac{\tilde{v}_{\rm gr}^2}{c^2}\right)^{-1/2} \! =\frac{\tom}{\tom_{\rm p}}\gg 1.
\eeq
All models considered below satisfy $\tom\gg \tom_{\rm p}$, so $\tilde v_{\rm gr}\approx c$.

Throughout this paper, we set $\tilde{v}_{\rm gr}=c$ in the fluid frame, which also implies $v_{\rm gr}=c$ in the lab frame. In particular, bulk acceleration and compression of the plasma exposed to the FRB is  calculated assuming that the FRB propagates with speed $c$, so the radio wave packet is static in the coordinate $\xi=t-z/c$. This approximation requires a more demanding condition than $\tg_{\rm gr}\gg 1$. Note that the plasma interacting with the radio wave packet forms a smooth compressive MHD wave of width $cT$. In MHD, compression tends to be redistributed across the fluid with an effective sound speed $\tilde v_{\rm s}$, which approaches $c$ in a magnetically dominated plasma. The deviation of FRB speed from $c$ may be  neglected in the MHD calculations of plasma motion in the radio wave packet if
\beq 
  \tg_{\rm gr}\gg\tg_{\rm s}=(1-\tilde v_{\rm s}^2/c^2)^{-1/2}.
\eeq
Then, the approximation $\tilde{v}_{\rm gr}=c$ can be used.

The value of $\tg_{\rm s}$ for a relativistic, hot, magnetized fluid is derived in Appendix~A in \cite{Beloborodov24}. It may be approximately described as
\beq
\label{eq:gs}
    \tg_{\rm s}\approx \sqrt{1+\sigma_{\rm eff}}, \qquad \sigma_{\rm eff}=\frac{\sigma}{\tg}=\frac{\tB^2}{4\pi \tg \me c^2\tn}.
\eeq
This expression for $\tg_{\rm s}$ becomes exact for a cold plasma ($\tg=1$). For a hot plasma, \Eq~(\ref{eq:gs}) approximates the accurate $\tg_{\rm s}$ within a numerical factor, which is close to unity (and depends on the plasma equation of state). The condition $\tg_{\rm gr}\gg\tg_{\rm s}$ is significantly more demanding than $\tg_{\rm gr}\gg 1$ if $\sigma_{\rm eff}\gg 1$. Then, it can be stated as 
\beq
   \frac{\tg_{\rm gr}}{\tg_{\rm s}}\approx \frac{\tom}{\tomL}\gg 1, \qquad \tomL=\frac{e\tB}{\me c\tg}.
\eeq


\section{Particle motion in the radio wave}
\label{app:motion}

Consider a linearly polarized radio wave packet of the form introduced in \Eqs~(\ref{eq:Ew}) and (\ref{eq:aw}):
\beq
   \Ew=E_0(\xi)\sin(\omega\xi), \quad\; \aw=a_0(\xi)\cos(\omega\xi),
\eeq 
where $\xi=t-z/c$. We choose $x$ and $y$ axes so that $\bEw$ oscillates along $x$, and $\bBw$ oscillates along $y$. The packet propagates along $z$ in a background field $\Bbg\ll E_0$ with $\bBbg\parallel\bBw$. 

The background magnetic field lines can drift along $z$ with some speed $\bD=\Ebg/\Bbg$. This speed describes the plasma bulk motion in the MHD picture. One could choose a reference frame so that the plasma ahead of the wave packet is at rest, $\bD=0$. However, $\bD$ can slowly vary with $\xi$ inside the packet (slowly compared  to the wave oscillation and the particle Larmor rotation). Then, there is no global inertial frame that would satisfy $\bD(\xi)=0$ across the packet. At each $\xi$, one can  define the local fluid frame $\tKF$ where $\tbD=0$ and $\tilde{E}_{\rm bg}=0$.

\subsection{Regular oscillation regime}
\label{stream}

In the regular oscillation regime, the plasma exposed to the radio wave moves as a coherent stream, i.e. the plasma particles have a common four-velocity $u^\mu$ varying as a function of the oscillation phase in the wave $\ph\equiv\omega\xi$. The four-velocity oscillates with frequency $\omega$ around the mean value $\langle u^\mu\rangle$, where the local averaging (over one oscillation) is defined as
\beq
\label{eq:averaging}
  \langle ...\rangle\equiv \frac{1}{2\pi}\oint ...\,d\ph,
\eeq
with the integral taken over $\ph= \omega\xi$ changing by $2\pi$. This averaging operation is equivalent to spatial averaging over $z$ at a fixed time $t$, since $d\ph=-\omega dz/c$ at $t=const$. 
In the fluid frame $\tKF$, $\langle \tilde u^i\rangle=0$ ($i=x,y,z$), so 
\beq 
  \langle \tilde u^\mu\rangle=(\langle\tilde \gamma\rangle,0,0,0). 
\eeq
Here, we assumed that the wave amplitude in the packet rises on a timescale $\tilde\xi_{\rm rise}\gg a_0\tomB^{-1}$; then, gyration of $\langle \tilde u^\mu\rangle$ in $\tBbg$ is not excited, as shown in \citetalias{Beloborodov22}. 

In an arbitrary frame $\KF$ moving along $z$, we have
\beq
\label{eq:u_av}
  \langle u^\mu\rangle=(\gav,0,0,\uzav).
\eeq 
It is related to $\langle\tilde{u}^\mu\rangle$ by Lorentz transformation, which gives (using $\tuzav=0$):
\beq
\label{eq:transf}
   \gav=\gD\tgav, \qquad \uzav=\uD\tgav, 
\eeq
where $\uD=\gD\bD$ and $\gD=(1-\bD^2)^{-1/2}$.
Note that in this appendix $\gamma$ denotes the instantaneous value of the particle Lorentz factor in the wave; in the main text, $\gamma$ stands for  $\langle\gamma\rangle$, as we omit $\langle...\rangle$ to simplify the notation. 

The oscillation form of $u^\mu(\ph)$  can be found from \Eq~(\ref{eq:dyn}) and the corresponding energy equation,
\beq
\label{eq:energy}
  \me c\,\frac{d\gamma}{dt}=e\bE\cdot\bb=e(\Ebg+E_{\rm wave})\beta_x.
\eeq
Since the problem is translationally invariant along $x$, the generalized momentum $mc u_x+eA_x/c$ is conserved, where $A_x=A_{\rm wave}+A_{\rm bg}$ and $\Bbg=dA_{\rm bg}/dz$. The small $\Bbg\ll E_0$ implies a small variation of $A_{\rm bg}$ over one wavelength, so  $u_x+a_{\rm wave}\approx const=\langle u_x\rangle =0$ and
\beq
\label{eq:uxstream}
  u_x(\ph) = -a_0\cos\ph.
\eeq
Conservation of generalized momentum $\me c u_y+eA_y/c$ (implied by $\partial_y=0$) gives $u_y(\ph)=0$ since $A_y=0$. 

The remaining component $u_z(\ph)$ and the Lorentz factor $\gamma(\ph)$ can be found by considering $u_\xi\equiv d\xi/d\tau=\gamma-u_z$, where $\tau$ is the proper time along the particle worldline. Subtracting the $z$-component of \Eq~(\ref{eq:dyn})  from \Eq~(\ref{eq:energy}), one finds
\beq
  \frac{du_\xi}{dt}=
  \frac{e(E\beta_x - B\beta_x)}{\me c} = \frac{e(\Ebg-\Bbg)\beta_x}{\me c}.
\eeq
Since $\Ebg\ll E_0$ and $\Bbg\ll E_0$, one can see that $u_\xi$ weakly varies during one wave oscillation, $u_\xi\approx const$. This fact implies that the proper density of the plasma stream $\hat\rho\approx const$ during each oscillation. Indeed, plasma density $\rho(\ph)$ satisfies the continuity equation
\beq
   \partial_t\rho+\partial_z(\rho v_z)=0 \quad \Rightarrow \quad
   \frac{d}{d\xi}\left[\rho(1-\beta_z)\right]=0.
\eeq
Hence, the proper density of the oscillating stream $\hat\rho=\rho/\gamma$ satisfies 
\beq
  \hat\rho u_\xi=const, 
\eeq
and so $\hat\rho\approx const$ during one oscillation. The weak variation of $u_\xi$ also implies the relation 
\beq
  u_\xi \approx \langle u_\xi\rangle =\langle\gamma-u_z\rangle=\frac{\tgav}{\c}, 
     \qquad \c\equiv \gD+\uD,
\eeq
where we used \Eq~(\ref{eq:transf}). 

In particular, in the fluid rest frame, $\tilde\c=1$ and $\tu_\xi=\tgav$. The value of $\tgav$ can be determined using the identity $\gamma^2=1+u_x^2+u_z^2$, which gives
\beq
  (u_z+u_\xi)^2=1+u_x^2+u_z^2
  \;\; \Rightarrow \;\, 2u_\xi\uzav+u_\xi^2=1+a^2,
\eeq
where we used $\langle u_x^2\rangle=\langle a_{\rm wave}^2\rangle= a^2$. In the fluid frame $\tKF$, $\tuzav=0$ and $\tu_\xi=\tgav$, so the above identity gives
\beq
\label{eq:tgav}
   \tgav^2=1+a^2.
\eeq

Explicit expressions for  $\gamma(\ph)$ and $u_z(\ph)$ can be obtained as follows. Using $dt=(\gamma/u_\xi)d\xi$, one finds from \Eq~(\ref{eq:energy})
\beq
\label{eq:g1}
   \frac{d\gamma}{d\xi}=\frac{e\Ew u_x}{\me c u_\xi} \;\;\Rightarrow\;  
   \gamma=\frac{a_0^2\cos 2\ph}{4u_\xi} + \gav,
\eeq
where $\Ew u_x=-E_0 a_0 \sin\ph \cos\ph$ and we used $u_\xi\approx const$ during one oscillation. \Eq~(\ref{eq:g1}) also determines  $u_z=\gamma-u_\xi$. Thus, one finds
\begin{align}
\label{eq:gstream}
    \gamma(\ph) &=\frac{\c a_0^2\cos 2\ph}{4\tgav} + \gD\tgav, \\
\label{eq:uzstream}
    u_z(\ph) &=\frac{\c a_0^2\cos 2\ph}{4 \tgav} + \uD\tgav.
\end{align}
These expressions hold in any frame $\KF$ moving along $z$. 

In particular, in the fluid frame, $\tilde{u}_{\rm D}=0$, $\tilde{\gamma}_{\rm D}=1$, and $\tilde\c=1$. In this frame, each particle executes a closed periodic orbit in the oscillating wave. Indeed, the net displacement in $\tilde z$ during one oscillation period is
 \beq
   \Delta \tilde z = \oint \tilde{v}_z d\tilde{t} = \oint \frac{\tilde u_z \, d\ph}{\tilde \omega \tilde u_\xi}= \frac{2\pi \tuzav}{\tilde \omega \tilde u_\xi}=0.
\eeq
Note that $\tBbg$ does not enter the solution for $\tilde{u}^\mu(\ph)$; it is the same as at $\tBbg=0$. This solution is well known: the particle executes an ``8''-shaped orbit \citep{Landau75}. The above derivation shows that the particles execute the same motion in a slowly varying $\Bbg$, as long as their motion is viewed in the local fluid frame $\tKF$. The role of $\Bbg$ and $\Ebg$ is that they control the fluid speed $\bD$ relative to a fixed lab frame: $\bD=\Ebg/\Bbg$. A slow (adiabatic) variation of $\bD$ changes the frame $\tKF$ while the shape of the particle orbit viewed in $\tKF$ remains unchanged. The size of the orbit and the average Lorentz factor $\tgav$ are determined by the local wave strength $a=\langle a_{\rm wave}^2\rangle^{1/2}$ (\Eq~\ref{eq:tgav}).

\subsection{Stochastic heating regime}
\label{stochastic_orbit}

In the regime of stochastic heating, the plasma particles gain very high energies. The particle motion viewed in the fluid frame $\tKF$ has two important features \citepalias{Beloborodov22}:
\\
(1) The particle gyrates with a Lorentz factor $\tg\gg a$ and Larmor frequency $\tomL=\tomB/\tg\ll \tom$. The gyration occurs in the magnetic field $B$ averaged over the fast wave oscillation: $\langle \Bbg+\Bw\rangle=\Bbg$. 
The particle oscillation in the radio wave creates a small modulation $|\Delta u|\sim a$ superimposed on the circular gyration. 
\\
(2) The particle experiences a resonance with the radio wave every Larmor rotation. It occurs at the gyro-phase $\psi$ where the particle velocity aligns with the wave propagation direction (i.e. with the $z$-axis); we define this gyro-phase as $\psi=0$. The resonance occurs in a narrow interval $\delta\psi\ll 1$ and suddenly changes $\tg$ by a random $\Delta\tg$ given in \Eq~(\ref{eq:Dtg}) below. 

Between the repeating resonances, the gyrating particle moves with $\tg\approx const$, i.e. its trajectory in the fluid frame is a circle in the $\tilde{x}\tilde{z}$ plane. Its four-velocity satisfies $\tu_x^2+\tu_z^2=\tg^2-1=const$ with
\beq
  \tu_x=\sqrt{\tg^2-1}\sin\psi, \qquad \tu_z=\sqrt{\tg^2-1}\cos\psi.
\eeq
Quantity $\tu_\xi(\psi)=\tg-\tu_z$ varies between $(2\tg)^{-1}$ and $2\tg$; its value averaged over one full Larmor orbit between two consecutive resonances equals $\tg$.
In a different frame $\KF$ moving along $z$, the fluid frame moves with speed $\bD$, and the particle four-velocity is determined by Lorentz transformation:
\beq
   u_x=\tu_x, \qquad u_z=\uD\tg+\gD\sqrt{\tg^2-1}\cos\psi.
\eeq

The stochastic pumping of $\tg$ by the repeating resonance events implies a quickly developing chaos, destroying the coherent motion of the plasma stream in the wave \citepalias{Beloborodov22}. As a result, at a given location $\xi$ in the radio wave packet, individual particles have random gyration phases $\psi$ and a broad distribution of $\tg$. The ensemble average $\langle\tg\rangle_{\rm ens}$ grows with time,  which means heating of the plasma.
It occurs when\footnote{\citetalias{Beloborodov22} defined the critical $\tg_\star$ slightly differently, using $a_0$ instead of $a=a_0/\sqrt{2}$. All quantities in \citetalias{Beloborodov22} were measured in frame $\tKF$ (where the background field is static)
and denoted without tilde.}
\beq
   \tg\gtrsim \tg_\star=\left(\frac{a^3}{b}\right)^{1/2} \qquad (a\gg 1),
\eeq
where $b=\tomB/\tom$. This condition implies that regular oscillations in the wave with $\tg\sim a\gg 1$ switch to the stochastic heating regime if $b\gtrsim a$ (the more accurate condition is $b>a/3$, see Figure~\ref{fig:trans}). The stochastic heating of gyrating particles consumes energy of the radio wave, i.e. it is an absorption process.

This process is controlled by particle energy exchange with the radio wave at each resonance, $ \Delta\tg$. It may be written in the form 
\beq
\label{eq:Dtg}
   \Delta\tg = H\left(\frac{a_0^3\tg}{b}\right)^{1/3}\cos\ph+\langle\Delta\tg\rangle,
\eeq 
where $H\approx 2.6$ is a numerical factor, $\ph$ is the (random) phase of the radio wave at the particle location when it has the gyro-phase $\psi=0$. The term $\propto\cos\ph$ describes random kicks to $\tg$; it is derived in \citetalias{Beloborodov22}. The term $\langle\Delta\tg\rangle$ is a small systematic gain (seen in Figure~2 in B22 as a small shift of the measured energy exchange events above the $\cos\ph$ curve). It was derived for particles with $\tg\gg a$ by \cite{Lyubarsky18}:
\beq
   \langle\Delta\tg\rangle = \frac{2^{7/3}\pi^2 a_0^2}{3^{5/3}[\Gamma(1/3)]^2\, b^{2/3}\tg^{1/3}}
    \approx 1.1\,\frac{a_0^2}{b^{2/3}\tg^{1/3}},
\eeq
where $\Gamma(1/3)\approx 2.68$. This result may be interpreted as usual synchrotron absorption of a weak monochromatic wave by particles with effective rest mass $m_{\rm eff}=\me \sqrt{1+a^2}$ gyrating with Lorentz factor $\tg_{\rm eff}=\tg/\sqrt{1+a^2}$ \citep{Lyubarsky18}. One can see that $\langle\Delta\tg\rangle$ is smaller than the typical random part of $\Delta\tg$ by the factor of $\sim (\tg_\star/\tg)^{2/3}$.

Particles with a given $\tg$ on average gain energy from the wave with rate $\dot\tg=\langle\Delta\tg\rangle\, \tomL/2\pi$. In principle, the net heating effect could be found by averaging the rate $\dot\tg$ over the distribution function $f(\tg)$: $d\langle\tg\rangle_{\rm ens}/dt=\int \dot\tg\,f(\tg)\,d\tg$. However, this approach faces two challenges. (1) One needs to know the evolving $f(\tg)$ shaped by particle diffusion in energy, which involves large random steps $\Delta\tg$, up and down. Thus, the stochastic nature of heating is essential. (2) $\dot\tg$ was derived in the limit of $\tg\gg a$ while a significant contribution to the integral comes from moderate $\tg\gtrsim a$ for the relevant distribution function.

In section~\ref{stochastic_heating}, we calculate the heating rate $d\langle\tg\rangle_{\rm ens}/dt$ using the method of \citetalias{Beloborodov22}: we follow the motion of a large ensemble of (initially static) particles that become exposed to the incoming radio wave, and find the evolution of $\langle\tg\rangle_{\rm ens}$. The result is described by \Eq~(\ref{eq:stochastic}) with $\chi\approx 0.8$.

\section{Plasma stress-energy tensor}
\label{app:T}

\subsection{Regular oscillation regime}

The coherently oscillating plasma stream has the proper density $\hat\rho=\rho/\gamma$ nearly constant during each oscillation (Appendix~\ref{stream}). Therefore, the plasma density averaged over one oscillation may be expressed as 
\beq
  \rhoav=\gav\, \hat\rho, 
\eeq
where the averaging operation $\langle...\rangle$ is defined in \Eq~(\ref{eq:averaging}). The fluid rest frame $\tKF$ satisfies $\tuzav=0$ and $\tgav=(1+a^2)^{1/2}$ (\Eq~\ref{eq:tgav}), so
\beq
  \trhoav=\sqrt{1+a^2}\, \hat\rho.
\eeq
Lorentz transformation of four-velocity gives $\gamma=\gD\tg+\uD \tu_z$, which implies $\gav=\gD\tgav$ and
\beq
   \rhoav=\gD\trhoav.
\eeq

The stress-energy tensor of the coherent oscillating stream is 
\beq
   T_{\rm stream}^{\mu\nu}(\ph)=\hat\rho \, u^\mu u^\nu.
\eeq
In the plane-parallel problem, the relevant components of $T^{\mu\nu}$ are $T^{tt}$, $T^{tz}$, and $T^{zz}$. They are determined by $u^t=\gamma$ and $u^z$  (\Eqs~(\ref{eq:gstream}) and (\ref{eq:uzstream})). The plasma stress-energy tensor averaged over one oscillation is
\beq
\label{eq:Tav}
  T_{\rm p}^{\mu\nu}=\langle T_{\rm stream}^{\mu\nu}\rangle=\frac{\hat\rho c^2}{2\pi}\oint  u^\mu u^\nu\,d\ph.
\eeq
The averaging operation is Lorentz invariant among all frames moving along $z$, since $\ph=inv$. As a result, \Eq~(\ref{eq:Tav}) yields a well-defined tensor
\beq
\label{eq:Tstream_av}
   \Tp^{\mu\nu} =\frac{c^2\trhoav \c^2 a^4}{8(1+a^2)^{3/2}}
   + c^2 \trhoav (1+a^2)^{1/2} \uD^\mu\uD^\nu,
\eeq
where $\uD^\mu=(\gD,\uD)$. It is straightforward to verify that the components $\Tp^{\mu\nu}$ with $\mu,\nu\in\{t,z\}$ satisfy the tensor transformation law under Lorentz boosts along $z$.\footnote{In the fluid frame, the components have the form $\Tp^{\tilde{\mu}\tilde{\nu}}=K+G\delta^{\tilde{\mu}}_0\delta^{\tilde{\nu}}_0$ where $K$ and $G$ are scalars. Tensor transformation with Lorentz matrix $\Lambda^\mu_{\tilde\mu}$ then gives $\Tp^{\mu\nu}=\Lambda^\mu_{\tilde\mu}\Lambda^\nu_{\tilde \nu} \Tp^{\tilde{\mu}\tilde{\nu}}=\c^2 K+G\uD^\mu\uD^\nu$, consistent with \Eq~(\ref{eq:Tstream_av}).}

In the main text, we drop $\langle...\rangle$ to simplify notation:
\beq
\nonumber
   \tgav\rightarrow\tg, \qquad \trhoav\rightarrow\trho.
\eeq

\subsection{Stochastic heating regime}

In the stochastic heating regime, the particle motion is dominated by gyration about $\bBbg$ with $\tg\gg a$. Fast oscillations in the wave occur with a smaller four-velocity  $|\Delta \bu|\sim a$ and are superimposed on the circular Larmor orbit. In this regime, the gyrating plasma may be approximated as a two-dimensional relativistic gas. Its internal energy density $\Up$ and pressure $\Pp$ (measured in frame $\tKF$) are given by
\beq
\label{eq:Up}
   \Up=2\Pp=(\tg-1)\trho c^2.  
\eeq 
The plasma stress-energy tensor is 
\beq
\label{eq:Tp}
  T^{\mu\nu}_{\rm p}=(\trho c^2+\Up+\Pp)\uD^\mu \uD^\nu+\eta^{\mu\nu}\Pp, 
\eeq
where $\eta^{\mu\nu}$ is the Minkowski metric.

\section{Radiative losses}
\label{app:losses}

A relativistic electron (or positron) with an instantaneous four-velocity $u^\mu=(\gamma,\bu)$ exposed to fields $\bE$ and $\bB$ radiates energy with rate \citep{Landau75}
\beq
\label{eq:dEe_general}
  \dEe=\frac{c\sT}{4\pi}\left[(\gamma\bE+\bu\times\bB)^2-(\bu\cdot\bE)^2\right],
\eeq
where $\sT$ is the Thomson cross section. $\dEe$ is a Lorentz-invariant quantity. The particle also radiates momentum with rate $\dot{\boldsymbol{\cal P}}_e=\bb\dEe/c$.

In our case, $\bE=\bEbg+\bEw$ is along $x$, $\bB=\bBbg+\bBw$ is along $y$, and $u_y=0$. Then, \Eq~(\ref{eq:dEe_general}) gives
\beq
\label{eq:dEe_wave1}
  \dEe=\frac{c\sT}{4\pi}\left\{\left[u_\xi E+\gamma (\Bbg-\Ebg)\right]^2-B^2+E^2\right\}.
\eeq
In both regimes, with regular oscillations or stochastic heating, emission in the fluid frame $\tKF$ occurs with $\tu_\xi\sim\tg\gg 1$. Then, for radio waves with amplitudes $\tE_0\gg\tBbg$, \Eq~(\ref{eq:dEe_wave1}) simplifies to 
\beq
\label{eq:dEe}
   \dEe = \frac{c\sT}{4\pi} u_\xi^2\Ew^ 2\left[ 1+{\cal O}\left(\frac{\tBbg}{\tilde E_{\rm wave}}\right)\right].
\eeq

Let $\delta t$ be the time it takes the particle to move one oscillation period $\delta\xi=2\pi/\omega$ through the wave:
\beq
   \delta t=\int dt = \oint \frac{\gamma \,d\ph}{u_\xi\omega}\approx \frac{2\pi}{\omega u_\xi}\gav,
\eeq
where we used the fact that quantity $u_\xi=\gamma-u_z$ weakly varies during one oscillation of the wave: $u_\xi\approx const$ (Appendix~\ref{stream}). The energy and momentum emitted by the particle during time $\delta t$ are 
\beq
   \delta\E_e= \! \frac{c\sT}{4\pi}\! \int \Ew^2\,u_\xi^2\,dt
   = \frac{c\sT u_\xi}{4\pi \omega} \oint \Ew^2\gamma\, d\ph,
\eeq
\beq
   \delta\boldsymbol{\P}_e=\frac{\sT}{4\pi} \int \Ew^2u_\xi^2 \bb dt
   = \frac{\sT u_\xi}{4\pi \omega}\! \oint \Ew^2 \bu\, d\ph.
\eeq
Thus, the emitted four-momentum $\delta \P_e^\mu=(\delta\E_e/c,\delta \boldsymbol{\P}_e)$ is
\beq
\label{eq:dPe4}
  \delta\P^\mu_e=\frac{2\sT U u_\xi}{\omega} \oint u^\mu \sin^2\!\ph\, d\ph,
\eeq
where $U=E_0^2/8\pi$ is the energy density of the radio wave. The average emission rate during $\delta t$ is
\beq
\label{eq:dPe4_dt}
  \frac{\delta\P^\mu_e}{\delta t}
  = \frac{2\sT U u_\xi^2}{\gav}
  \langle u^\mu \sin^2\!\ph\rangle.
\eeq

\subsection{Regular oscillation regime}

In the regular oscillation regime, the components of $u^\mu(\ph)$ are given by \Eqs~(\ref{eq:uxstream}), (\ref{eq:gstream}), and (\ref{eq:uzstream}), with $\gav=\gD\sqrt{1+a^2}$ and $u_\xi=\c^{-1}\sqrt{1+a^2}$. Then, \Eq~(\ref{eq:dPe4_dt}) yields two non-zero components of $\delta\P^\mu_e/\delta t$: 
\begin{align}
\label{eq:dE_dt_stream}
   \frac{\delta\E_e}{\delta t} &=  \frac{c\sT U }{\c^2}
 \left[ 1+a^2 -\frac{\c a^2}{4\gD}\right], \\
\label{eq:dP_dt_stream}
   \frac{\delta\P^z_e}{\delta t} &=  \frac{\sT U }{\c^2}
 \left[ \bD(1+a^2) -\frac{\c a^2}{4\gD}\right].
\end{align}
These expressions hold in any frame moving along $z$. While the instantaneous losses $\dEe$ are Lorentz-invariant, their average value is not --- $\delta\E_e/\delta t$ depends on the frame (because of the term $-\c a^2/4\gD$). Note that in the fluid frame $\tKF$ the particle radiates $z$-momentum with rate $\delta\tilde\P^z_e/\delta \tilde t = -\sT \tilde U a^2/4\neq 0$, even though its orbit has the symmetric ``8'' shape.

\subsection{Stochastic heating regime}

In the regime of stochastic heating, the particle executes gyration in $\Bbg$ with four-velocity $u^\mu$ weakly modulated by the fast oscillations in the wave. Therefore,  in \Eq~(\ref{eq:dPe4}) one can take $u^\mu\approx const$, and the emitted four-momentum averaged over one wave oscillation becomes 
\beq
\label{eq:dPe4_dt_}
  \frac{\delta \P^\mu_e}{\delta t}
  = \sT U\, \frac{ u_\xi^2 u^\mu}{\gamma}.
\eeq

In the MHD description of plasma dynamics, the radiative losses averaged over the fast oscillation in the wave should be further averaged over Larmor rotation. This is convenient to do in the fluid frame $\tKF$ because in this frame the particle motion between the successive resonances has $\tg\approx const$ (section~\ref{stochastic_orbit}), so its orbit is a Larmor circle with the orbital time
\beq
  \Delta\tilde t=\frac{2\pi\tg}{\tomB}\gg \delta \tilde t.
\eeq
The gyro-phase $\psi$ is defined as the angle between the particle velocity $\tilde\bb$ and the $z$-axis, so 
\beq 
  \tilde\beta_z=\tilde\beta\cos\psi\approx \cos\psi. 
\eeq
Integration over $\Delta \tilde t$ can be reduced to the integration over $0<\psi<2\pi$ using $d\psi/d\tilde t=\tomL=\tomB/\tg$. Then, the energy and $z$-momentum emitted by the particle during $\Delta \tilde t$ can be expressed as 
\beq
  \Delta \tilde\E_e = \frac{c \sT \tU}{\tomL} \! \int_0^{2\pi} \!\! \tu_\xi^2\,d\psi,
\eeq
\beq
\label{eq:dP_stochastic}
  \Delta \tilde\P_e  =  \frac{\sT \tU}{\tomL} \! \int_0^{2\pi} \!\! \tu_\xi^2\tilde\beta_z\, d\psi. 
\eeq
Substituting $\tu_\xi=\tg(1-\tbeta_z)$ and $\tbeta_z\approx \cos\psi$, we find
\beq
\label{eq:dE_stochastic}
  \frac{\Delta \tilde\E_e}{\Delta\tilde t} =\frac{3}{2} c \sT \tU\tg^2,
  \qquad  
  \frac{\Delta \tilde\P_e}{\Delta\tilde t} = -\sT \tU\tg^2. 
\eeq
Although the particle orbit is nearly circular, it radiates a large net momentum along $-z$. This occurs because the instantaneous particle acceleration is dominated by the fast-oscillating  electromagnetic field of the wave (even though the modulation effect of the wave on the particle velocity is small, so that the particle continues to move on the circular Larmor orbit with $\tg\approx const$). The fact that the wave field propagates along $\tilde z$ breaks the emission symmetry $-\tilde z\leftrightarrow \tilde z$, as seen in \Eq~(\ref{eq:dP_stochastic}): $\tu_\xi^2=(\tg-\tu_z)^2$ decreases at $\tu_z>0$ and increases at $\tu_z<0$.

The average loss rate in another frame $\KF$ moving along $z$ is determined by the Lorentz transformation,
\begin{align}
  \Delta\E_e &= \gD\Delta\tilde\E_e+\uD c\Delta\tilde\P_e, \\
  c\Delta\P_e &= \uD\Delta\tilde\E_e+\gD c\Delta\tilde\P_e, 
\end{align}
and $\Delta t=\gD\Delta\tilde t$. Thus, we find 
\begin{align}
\label{eq:dE_dt_stochastic}
  \frac{\Delta \E_e}{\Delta t} &= \frac{c \sT \Uw}{\c^2}\left( \frac{3}{2}-\bD\right) \tg^2, \\
\label{eq:dP_dt_stochastic}
  \frac{\Delta \P_e}{\Delta t} &= \frac{\sT \Uw}{\c^2}\left( \frac{3}{2}\bD-1\right) \tg^2,
\end{align}
where we used $\tUw=\Uw/\c^2$, which expresses transformation of the wave energy density from lab frame $\KF$ to the fluid frame $\tKF$ (\Eq~\ref{eq:transform}).

 \newpage

\bibliography{ms}

\end{document}